\documentclass[11pt,a4paper]{article}
\pdfoutput=1
\usepackage{jheppub}

\usepackage{multirow, graphicx,amssymb,url,mathrsfs,amsmath}
\usepackage{wrapfig,boxedminipage,setspace,subfigure,epsfig}
\usepackage{amsxtra,amstext,latexsym,dsfont,amsfonts}
\usepackage{color,eucal}
\usepackage[dvipsnames]{xcolor}
\usepackage{float}
\usepackage{slashed}

\renewcommand{\d}{\delta}

\newcommand{\dd}{\mathrm{d}}

\newcommand{\grad}{\nabla}

\newcommand{\tb}{\mbox{${\tilde{k}}$}}
\newcommand{\bb}{\mbox{${\bar{k}}$}}
\newcommand{\bT}{\mbox{${\bar{T}}$}}

\newcommand{\tT}{\mbox{$\tilde{T}$}}

\newcommand{\tQ}{{\tilde{Q}}}
\newcommand{\tmu}{{\tilde{\mu}}}

\title{Homes' law in holographic superconductor with linear-$T$ resistivity }

\author[a,b]{Hyun-Sik Jeong}
\author[c]{and Keun-Young Kim}

\emailAdd{hyunsik@ucas.ac.cn}
\emailAdd{fortoe@gist.ac.kr}

\affiliation[a]{School of physics $\&$ CAS Center for Excellence in Topological Quantum Computation, University of Chinese Academy of Sciences, Zhongguancun east road 80, Beijing 100049, China}
\affiliation[b]{Kavli Institute for Theoretical Sciences, University of Chinese Academy of Sciences, \\ Zhongguancun east road 80, Beijing 100049, China}
\affiliation[c]{School of Physics and Chemistry, Gwangju Institute of Science and Technology, \\
123 Cheomdan-gwagiro, Gwangju 61005, Korea}

\abstract{
Homes' law, $\rho_{s} = C \, \sigma_{DC} \, T_{c}$, is a universal relation of superconductors between the superfluid density $\rho_{s}$ at zero temperature, the critical temperature $T_{c}$ and the electric DC conductivity $\sigma_{DC}$ at $T_c$.
Experimentally, Homes' law is observed in high $T_c$ superconductors with linear-$T$ resistivity in the normal phase, giving a material independent universal constant $C$.
By using holographic models related to the Gubser-Rocha model, we investigate how Homes' law can be realized together with linear-$T$ resistivity in the presence of momentum relaxation. 
We find that strong momentum relaxation plays an important role to exhibit Homes' law with linear-$T$ resistivity. 
}

\begin{document}

\maketitle

\section{Introduction}

Holographic methods (gauge/gravity duality) have been providing novel and effective ways to study universal properties of strongly correlated systems.
The representative examples would be the holographic lower bound of the ratio of shear viscosity to entropy density, linear-$T$ resistivity and Hall angle of strange metals~\cite{Hartnoll:2016apf,Zaanen:2015oix,Ammon:2015wua,Baggioli:2019rrs,Hartnoll:2009sz,Herzog:2009xv}.

In this paper, using holography, we study another universal property of strongly coupled systems, which is observed in high $T_{c}$ superconductors and some conventional superconductors: Homes' law~\cite{Homes:2005aa,Homes:2004wv}.
Homes' law is an empirical relation between the superfluid density at $T=0$ ($\rho_{s}(T=0)$), the phase transition temperature ($T_{c}$), and the electric DC conductivity in the normal phase close to $T_c$ ($\sigma_{DC}(T_{c})$):
\begin{align}
\rho_{s}(T=0) \,=\, C \, \sigma_{DC}(T_{c}) \, T_{c} \,, \label{HOMEHOME}
\end{align}
where $C$ is a material independent universal constant.
For instance, $C\sim4.4$ for ab-plane high $T_{c}$ superconductors and clean BCS superconductors or $C\sim8.1$ for c-axis  high $T_{c}$ superconductors and BCS superconductors in the dirty limit.

In order to study Homes' law in holography, first one may need to construct the holographic superconductor model.
Using the complex scalar field, the holographic superconductor model was originally proposed by Hartnoll, Herzog, and Horowitz~\cite{Hartnoll:2008vx,Hartnoll:2008kx} (the HHH model). Thereafter, there has been extensive development and extension of the HHH model in \cite{Hartnoll:2009sz,Herzog:2009xv,Horowitz:2010gk,Cai:2015cya}. For the recent development of holographic superconductors, see also \cite{Kim:2013oba,Gouteraux:2019kuy,Gouteraux:2020asq,Arean:2021brz,Donos:2021pkk,Ammon:2021slb} and references therein.

Since the HHH model is a translational invariant theory, $\sigma_{DC}$ is infinite so  $C$ in \eqref{HOMEHOME} is not well defined.
Thus, in order to investigate Homes' law, one may need to break the translational invariance to render $\sigma_{DC}$ finite. 
In holography, there are several methods to incorporate momentum relaxation and yield a finite $\sigma_{DC}$. For instance, the bulk fields in gravity with the inhomogeneous boundary conditions~\cite{Horowitz:2012ky}, massive gravity models~\cite{Vegh:2013sk}, Q-lattice models~\cite{Donos:2013eha}, the linear axion model~\cite{Andrade:2013gsa}, and the helical lattice model with a Bianchi VII$_0$ symmetry~\cite{Donos:2012js}.
Using these models, holographic superconductors in the presence of the momentum relaxation have been investigated in \cite{Erdmenger:2015qqa,Horowitz:2013jaa,Zeng:2014uoa,Ling:2014laa,Andrade:2014xca,Kim:2015dna,Baggioli:2015zoa,Baggioli:2015dwa,Kim:2016hzi,Kim:2016jjk,Ling:2016lis,Jeong:2018tua}.

In the aforementioned holographic superconductor models with momentum relaxation, Homes' law has been studied only in several models~\cite{Erdmenger:2015qqa,Kim:2016jjk,Kim:2016hzi}.\footnote{See \cite{Erdmenger:2012ik} for an early attempt to study Homes' law in holography without momentum relaxation. See also \cite{Ling:2016lis} for a modified version of Homes' law with Weyl corrections.}
For those models, there are parameters for the strength of momentum relaxation, which may specify material properties. Thus, in the holographic setup, Homes' law means that $C$ in \eqref{HOMEHOME} is constant independent of momentum relaxation parameters.  
In \cite{Erdmenger:2015qqa}, using \textit{the helical lattice model}, Homes' law was studied with the amplitude and the pitch of the helix as momentum relaxation parameters.
In \cite{Kim:2016hzi}, \textit{the linear axion model} was studied for Homes' law with the the proportionality constant to spatial coordinate, $k$ in \eqref{3charge}, for the strength of momentum relaxation.\footnote{With the linear axion model, the normal phase has been studied in \cite{Kim:2015dna,Kim:2014bza,Kim:2015sma,Kim:2015wba}, the superconducting phase in \cite{Andrade:2014xca,Kim:2015dna}, and the fermionic phase in \cite{Jeong:2019zab}. See also \cite{Baggioli:2021xuv} for the review/recent development of the holographic axion model.}
In \cite{Kim:2016jjk}, \textit{the Q-lattice model} was used to study Homes' law with the lattice amplitude/wavenumber for momentum relaxation parameter.

In all holographic studies so far, Homes' law has not been  well realized in that, in \cite{Erdmenger:2015qqa,Kim:2016jjk}, Homes' law is satisfied only for some restricted parameter regime in which underlying physics has not been clearly understood yet or Homes' law is not simply satisfied in \cite{Kim:2016hzi}.
Therefore, the fundamental understanding and the physical mechanism of Homes' law is still lacking and it would be important to study Homes' law with other holographic models.

In this paper, we study Homes' law in the holographic superconductor model based on \textit{the Gubser-Rocha model}~\cite{Gubser:2009qt} with the axion field to have momentum relaxation~\cite{Davison:2013txa,Zhou:2015qui,Kim:2017dgz,Jeong:2018tua,Liu:2021qmt}.\footnote{When we remove a dilaton field in the Gubser-Rocha model, it becomes the linear axion model in \cite{Kim:2016hzi}.}
Our main motivation to choose this model is that it exhibits linear-$T$ resistivity in its normal phase~\cite{Davison:2013txa,Zhou:2015qui,Jeong:2018tua}. 
In particular, in \cite{Jeong:2018tua}, it was shown that the linear-$T$ resistivity is robust above $T_c$ in the {\it strong} momentum relaxation limit, which is similar to the experimental result for normal phases (strange metal phases) of high $T_c$ superconductors. 
{We will examine if Homes' law can appear also in the strong momentum relaxation limit and also study its relation with the linear-$T$ resistivity.}

The property of linear-$T$ resistivity is important for two reasons. First, it is another universal property in the normal phase of high $T_c$ superconductors, so it is in fact a necessary property even before discussing Homes' law. Second, it has been proposed that the Homes' law can be explained by considering the Planckian dissipation \cite{Zaanen:2004aa}, which is related with the linear-$T$ resistivity. 
Therefore, the linear-$T$ resistivity can be a key to understand the physics of Homes' law. Because there has been no holographic model studying Homes' law together with linear-$T$ resistivity\footnote{Note that other holographic studies for Homes' law~\cite{Erdmenger:2015qqa,Kim:2016jjk,Kim:2016hzi} did not show the linear-$T$ resistivitiy.} our study is a necessary and important step to investigate Homes' law.
Moreover, the Gubser-Rocha model with the axion field allows an analytic solution so that more tractable analysis is available for the normal phase. Note that most holographic studies in \cite{Erdmenger:2015qqa,Kim:2016jjk,Kim:2016hzi} do not allow analytic solution for normal phase so one needs to resort to numerical methods.

 
As one of the ingredient of our holographic superconductor model,  inspired by \cite{Lucas:2014zea,Cremonini:2016bqw}, we introduce the non-trivial coupling, $B(\phi)$, between the dilaton field $\phi$ and the complex scalar field $\Phi$ for condenstate, and study the role of the coupling in Homes' law. In \cite{Lucas:2014zea,Cremonini:2016bqw}, using the scaling property from $B(\phi)$, the superconducting instabilities have been investigated in which the translational invariance was not broken. Thus, our work might be considered as its generalization with momentum relaxation. Note that $B(\phi)$ was taken to be a mass term of $\Phi$, $B(\phi)=M^2$, in the previous literature for Homes' law~\cite{Erdmenger:2015qqa,Kim:2016jjk,Kim:2016hzi}. We find that Homes' law may not be realized with this trivial mass term.

 

This paper is organized as follows. 
In section \ref{section2modelex}, we introduce the holographic superconductor models based on the Gubser-Rocha model with the axion fields. In normal phase, we review how to obtain the linear-$T$ resistivity analytically. For superconductor phase, we introduce the coupling term $B(\phi)$ and review its properties.  We also study superconducting instability with $T_c$.
In section \ref{sec3}, we numerically compute the optical conductivity and study the superfluid density. Using the linear-$T$ resistivity in section \ref{section2modelex} with the superfluid density in section \ref{sec3}, we study Homes' law. We also discuss the role of the coupling $B(\phi)$ for Homes' law. 
In section \ref{sec4}, we conclude.

%
\section{Superconductor based on the Gubser-Rocha model}\label{section2modelex}

\subsection{Model}
We study a holographic superconductor model based on Einstein-Maxwell-Dilaton-Axion theory:
\begin{align} \label{action1}
\begin{split}
S & =  S_1 + S_2+S_3  =  \int \dd^4x\sqrt{-g}\left( \mathcal{L}_1 + \mathcal{L}_2 + \mathcal{L}_3  \right) \,, \\ 
\mathcal{L}_1 &= R \,-\,\frac{1}{2}(\partial{\phi})^2 \,-\, \frac{1}{4} e^{\frac{\phi}{\sqrt{3}}}\,F^2 \,+\, 6\cosh \left(\frac{\phi}{\sqrt{3}}\right)  \,, \\
\mathcal{L}_2 &= -\frac{1}{2}\sum_{I=1}^{2}(\partial \psi_{I})^2 \,, \qquad \mathcal{L}_3 =  -|D\Phi|^2 \, -B(\phi) |\Phi|^2  \,,
\end{split}
\end{align}
where we set units such that the AdS radius $L=1$, and the gravitational constant $16\pi G = 1$.
The action \eqref{action1} consists of three actions.
The first action $S_1$ is the Einstein-Maxwell-Dilaton theory, which is called `Gubser-Rocha model'~\cite{Gubser:2009qt} composed of three fields: metric $g_{\mu\nu}$, a $U(1)$ gauge field $A_{\mu}$ with the field strength $F=\dd A$, and the scalar field $\phi$ so called `dilaton'.
The metric and gauge field are for a quantum field theory at finite temperature and density, while the dilaton field was originally introduced to make the vanishing entropy density ($s$) at zero temperature ($T$) as $s \sim T$~\cite{Gubser:2009qt,Davison:2013txa}.
The second action $S_2$ is added for the momentum relaxation: the `axion' field $\psi$ breaks the translational invariance so that the resistivity becomes finite~\cite{Andrade:2013gsa,Davison:2013txa,Zhou:2015qui,Kim:2017dgz,Jeong:2018tua}.\footnote{The axion-type model is related to the St\"uckelburg formulation of a massive gravity theory~\cite{Vegh:2013sk,Davison:2013jba,Blake:2013bqa,Blake:2013owa}.}
The third action $S_3$ is for the superconducting phase~\cite{Hartnoll:2008vx}, which is composed of a complex scalar field $\Phi$, the coupling $B(\phi)$, and the covariant derivative defined by $D_{\mu} := \grad_{\mu} -i q A_{\mu}$.

The action \eqref{action1} yields the equations of motion of matter fields
\begin{align}\label{MATTEREOM}
&\grad_{\mu}(e^{\frac{\phi}{\sqrt{3}}} F^{\mu\nu})-iq\Phi^{*}(\partial^{\nu}-iqA^{\nu})\Phi+iq\Phi(\partial^{\nu}+iqA^{\nu})\Phi^{*}=0  \,,  \\ \label{GFEOM}
&\grad^2\phi-\frac{1}{4\sqrt{3}}e^{\frac{\phi}{\sqrt{3}}} \, F^2 + 2\sqrt{3} \,\sinh \left(\frac{\phi}{\sqrt{3}}\right) -B'(\phi)|\Phi|^2 =0 \,, \\
&\grad^{2}\psi_{I}=0 \,, \\
& D^{2}\Phi-B(\phi)\Phi=0  \label{CSFEOM}\,,
\end{align}
and the Einstein's equation
\begin{equation}\label{EiensteinEq}
\begin{split}
&R_{\mu\nu} -\frac{1}{2}g_{\mu\nu}\left[R-\frac{1}{4} e^{\frac{\phi}{\sqrt{3}}} F^2 -\frac{1}{2}(\partial{\phi})^2+6\cosh \left(\frac{\phi}{\sqrt{3}}\right) -\frac{1}{2}\sum_{I=1}^{2}(\partial \psi_{I})^2 -|D\Phi|^2 -B(\phi)|\Phi|^2    \right] \\
& \qquad =\frac{1}{2}e^{\frac{\phi}{\sqrt{3}}} F_{\mu\d}F_{\nu}{^\d}+\frac{1}{2}\partial_{\mu}\phi \partial_{\nu}\phi+\frac{1} {2}\sum_{I=1}^{2}(\partial_{\mu}\psi_{I}\partial_{\nu}\psi_{I})+\frac{1}{2}\left(D_{\mu}\Phi D_{\nu}^{*}\Phi^{*}+D_{\nu}\Phi D_{\mu}^{*}\Phi^{*} \right) \,.
\end{split}
\end{equation}
%

%
\subsection{Linear-$T$ resistivity in strange metal phase: a quick review}
Let us first review the normal phase ($\Phi=0$), $S=S_1+S_2$, i.e., the Gubser-Rocha model with momentum relaxation. The purpose of this review is not only to organize this paper in a self-contained manner, but also collect useful results, linear-$T$ resistivity, to study our main objective, Homes' law, in section \ref{sec3}. We refer to \cite{Jeong:2018tua} for more detailed explanation of the normal phase.

In normal phase ($\Phi=0$), the analytic solution is available~\cite{Cremonini:2016bqw,Zhou:2015qui,Kim:2017dgz,Jeong:2018tua}:
\begin{equation}\label{3charge}
\begin{split}
&\dd s^2  = - f(r) \dd t^2+\frac{1}{f(r)}\dd r^2+h(r)(\dd x^2+\dd y^2),\\
&f= r^{1/2} (r+Q)^{3/2}\left(1 - \frac{k^2}{2(r+Q)^2} -\frac{(r_h+Q)^3}{(r+Q)^3} \left(1 - \frac{k^2}{2(r_{h}+Q)^2} \right) \right), \\ &h=r^{1/2} (r+Q)^{3/2},\\
&A_t=\sqrt{3Q(r_h+Q)\left(1-\frac{k^2}{2(r_{h} + Q)^2}\right)}\left(1-\frac{r_h+Q}{r+Q}\right),\quad \phi=\frac{\sqrt{3}}{2}\ln(1+Q/r) \, , \\
&\psi_{1}= k \, x \,, \quad \psi_{2}= k \, y \,,
\end{split}
\end{equation}
where $r_{h}$ denotes the horizon radius, $k$ controls a strength of the momentum relaxation, and $Q$ is a parameter which can be expressed with physical parameters: temperature ($T$), chemical potential ($\mu$) or momentum relaxation parameter ($k$).
The temperature and chemical potential reads
\begin{align}
&T= \frac{f'(r)}{4\pi}\Bigr|_{r_{h}} = {r_h}\frac{6(1+\tQ)^{2}-\tb^{2}}{8\pi(1+\tQ)^{3/2}} = r_h \tT \,,  \label{tTeq} \\
&\mu= A_t(\infty) = r_h \sqrt{3\tQ(1+\tQ)\left(1-\frac{\tb^{2}}{2(1+\tQ)^{2}}\right)}  = r_h \tmu  \label{tmueq}\,,
\end{align}
where
\begin{equation} \label{TILDE}
\tilde{Q} :=  \frac{Q}{r_{h}} \,, \qquad \tilde{k} :=\frac{k}{r_{h}} \,, \qquad \tilde{T} := \frac{T}{r_{h}} \,, \qquad \tilde{\mu} := \frac{\mu}{r_{h}} \,.
\end{equation}
Now one can obtain the dimensionless physical quantities at finite density, $T/\mu$ and $k/\mu$, as
\begin{align}
&\bT := \frac{T}{\mu} = \frac{\tT}{\tmu} =  \frac{6(1+\tQ)^{2}-\tilde{k}^{2}}{4\sqrt{6}\pi\sqrt{\tQ(1+\tQ)^{2}(2(1+\tQ)^{2}-\tilde{k}^{2})}} \,,  \label{bteq} \\ 
&\bb := \frac{k}{\mu} = \frac{\tb}{\tmu} =  \sqrt{\frac{2(1+\tQ)\tilde{k}^{2}}{3\tQ(2(1+\tQ)^2-\tilde{k}^2)}} \,, \label{bbeq}
\end{align}
where we used \eqref{tTeq}-\eqref{TILDE}.

\paragraph{The linear-$T$ resistivity:}
The electric DC conductivity at $\Phi=0$ can be obtained~\cite{Donos:2014cya,Blake:2013bqa,Kim:2015wba,Kim:2015sma} as
\begin{equation} \label{conduct1}
\sigma_{DC} := e^{\frac{\phi}{\sqrt{3}}} + \frac{A_{t}'^{2} \, h(r) \, e^{\frac{2\phi}{\sqrt{3}}}}{k^2}\Bigr|_{r\rightarrow r_{h}} =  \sqrt{1+\tQ} + \frac{\sqrt{1+\tQ}}{\bb^2} \,.
\end{equation}
Using \eqref{bteq}-\eqref{bbeq}, one can express $\tQ$ as a function of $\bT$ and $\bb$ analytically, i.e. $\tQ(\bT, \bb)$, this implies that the electric DC conductivity \eqref{conduct1} can also be expressed in terms of $\bT$ and $\bb$ as $\sigma_{DC}(\bT, \bb)$.

With the analytic expression of $\tQ(\bT, \bb)$, one can find that the Gubser-Rocha model can exhibit linear-$T$ resistivity, the resistivity ($\rho = 1/\sigma_{DC}$) is linear in temperature, for two cases\footnote{For more details, see \cite{Jeong:2018tua}.}
\begin{align}
&\sigma_{DC} \,\sim\, \frac{\sqrt{3}\left(1+\bb^2\right)^2}{2 \pi \bb^2 \sqrt{4+6 \bb^2}}\frac{1}{\bT}  \,, \qquad\quad   (\bT \ll 1 \,\,\text{for given} \,\, \bb) \,, \label{sigmasmallT} \\
&\sigma_{DC}   \,\sim\, \frac{\bb}{2\sqrt{2}\pi \,\bT}  \,, \qquad \qquad \qquad\,\,     \ \ (\bb \gg 1 \,\,\text{for given} \,\, \bT) \,,\label{sigmabigb} 
\end{align}
where $\tQ(\bT, \bb)$ has the asymptotic form as 
\begin{align}
&\tQ \sim \frac{3(1+\bb^2)^{2}}{8\pi^{2}(2+3\bb^2)\bT^2}  \,, \qquad\qquad\quad  (\bT \ll 1 \,\,\text{for given} \,\, \bb) \,, \label{tQ1}\\ 
&\tQ \sim \frac{\bb^2}{8\pi^2 \bT^2} \,, \qquad \qquad \qquad\qquad\quad  (\bb \gg 1 \,\,\text{for given} \,\, \bT) \,. \label{tQ3}
\end{align}

The former case ($\bT \ll 1$), \eqref{sigmasmallT}, is related to the result in \cite{Davison:2013txa} and this linear-$T$ resistivity is due to the fact that the Gubser-Rocha model has the Conformal to AdS$_2 \times R^{2}$ IR geometry~\cite{Gouteraux:2014hca}.\footnote{In the semi-locally critical limit where the dynamical exponent $z \rightarrow \infty$ and a hyperscaling violating exponent $\theta \rightarrow -\infty$ with the fixed $\theta/z = -\eta$, the Einstein-Maxwell-Dilaton-Axion theory has the  Conformal to AdS$_2 \times R^{2}$ IR geometry with the parameter $\eta$ and the resistivity behaves as $\rho\sim T^{\eta}$. The linear-T resistivity appears in Gubser-Rocha model because $\eta=1$ for Gubser-Rocha model.}
The other case ($\bb \gg 1$), \eqref{sigmabigb}, will be one of important ingredients of our main results for Homes' law in section \ref{sec3}.\footnote{Note that, as pointed it out in \cite{Jeong:2018tua,Ahn:2019lrh}, \eqref{sigmasmallT} may not guarantee that the linear-$T$ resistivity is robust up to high temperature. Thus, phenomenologically \eqref{sigmabigb} would be a more relevant condition to show the linear-$T$ resistivity above $T_c$, which is similar to experiments and checked in holography~\cite{Jeong:2018tua}.}

%
\subsection{Superconducting phase and the coupling $B(\phi)$}\label{sec23}

Let us study the superconducting phase based on the Gubser-Rocha model, $S=S_1+S_2+S_3$ \eqref{action1}, which will also be used for Homes' law in section \ref{sec3}.
Note that in order for the description of holographic superconductors, first we need to specify the form of the coupling $B(\phi)$. In this section \ref{sec23}, we first review how to introduce the coupling \eqref{COUPLING} chosen in this paper and study the superconducting instability with the critical temperature $T_c$.

Although we will examine Homes' law with the fully back-reacted background geometry in section \ref{sec3}, it would be instructive to treat a complex scalar field $\Phi$ as a perturbation field on top of the background geometry of Gubser-Rocha model \eqref{3charge}. 

There are three main reasons why we perform the perturbative (i.e., without back-reaction) analysis here. 
First, we can investigate the properties of the coupling $B(\phi)$ with the analytic IR scaling geometry. Moreover, one may also try to obtain the analytic instability condition.
Second, we may use the perturbative analysis as a guide for the study of Homes' law in next section, i.e., we will study $T_{c}$, one of the main ingredients for Homes' law, in the simple (i.e., no back-reaction) setup and show that $T_{c}$ from the perturbative analysis is consistent with $T_{c}$ in the presence of the back-reaction.
Third, our perturbative analysis for $T_c$ will be an extension to the previous work~\cite{Cremonini:2016bqw} where the translational symmetry was not broken ($k/\mu=0$).

Note that the analysis for $T_c$ would be important not only for Homes' law, but also to find the condition for high $T_c$ superconductors having linear-$T$ resistivity. As we will show, the trivial coupling, $B(\phi)=M^2$, would not be enough to have the superconducting phases at strong momentum relaxation limit, which is connected to the normal phase showing linear-$T$ resistivity \eqref{sigmabigb}.

\subsubsection{UV completion of $B(\phi)$}
Using the scaling properties in the IR region, one minimal way to choose the coupling $B(\phi)$ was introduced in \cite{Cremonini:2016bqw}. Here we not only review the method in \cite{Cremonini:2016bqw}, but also extend the analysis in \cite{Cremonini:2016bqw} to the case at finite $k/\mu$.
\paragraph{The extremal IR geometry:}
In order to investigate the IR scaling properties, we first need to have the extremal IR geometry which can be obtained from \eqref{3charge} in $T\rightarrow0$ limit. 
From \eqref{bteq}, one can find the condition for $T=0$ as $r_{h}/Q \rightarrow 0$ (or $\tQ \rightarrow \infty$). Note that there is another mathematical possibility to obtain $T=0$ from the relation between $\tQ$ and $\tilde{k}$ such that $6(1+\tQ)^{2}-\tilde{k}^{2} = 0$. However, this another condition can be ruled out for the physical reason: it gives the imaginary chemical potential \eqref{tmueq} and momentum relaxation \eqref{bbeq}.\footnote{Note also that $\tQ$ should be positive to be thermodynamically stable~\cite{Kim:2017dgz}.}

Then, using the condition $r_{h}/Q \rightarrow 0$ with the following coordinate transformation\footnote{There would be other coordinate transformation to express Conformal to AdS$_2 \times R^{2}$ geometry. For the comparison with the previous literature \cite{Cremonini:2016bqw}, we have used \eqref{ctCL}.} in \eqref{3charge}
\begin{equation}\label{ctCL}
\begin{split}
\rho=\sqrt{\frac{Q}{3r}}, \qquad t^{\prime} = \sqrt{1-\frac{k^2}{6 Q^2}} \,t \,,
\end{split}
\end{equation}
one can express the extremal IR geometry as
\begin{equation}\label{ir3charge}
\begin{split}
&\dd s^2  = \frac{Q^2}{\sqrt{3}}\frac{1}{\rho}\left[- \frac{\dd t^{\prime 2}}{\rho^2}+\frac{8}{6Q^2-k^2} \frac{\dd \rho^2}{\rho^2}+\dd x^2+\dd y^2\right],\\
&A_{t^{\prime}}=\sqrt{\frac{2Q^2-k^2}{6Q^2-k^2}}\frac{Q}{\rho^2},\quad\quad \phi=\sqrt{3}\ln(\sqrt{3}\rho) \, ,
\end{split}
\end{equation}
where the IR is located at $\rho\rightarrow\infty$. Note that \eqref{ir3charge} corresponds to Conformal to AdS$_2 \times R^{2}$ geometry and it is consistent with the one in \cite{Cremonini:2016bqw} at $k=0$.

%
\paragraph{The coupling $B(\phi)$ in IR:}
The complex scalar field Lagrangian in \eqref{action1}, $\mathcal{L}_3$, can be written as follows.
\begin{align} \label{EFFFFMASSS}
\begin{split}
-\mathcal{L}_3 = (\partial \Phi)^2 \,+\, \left[q^2 A^2 +B(\phi)\right] \Phi^2   \,,
\end{split}
\end{align}
where $\Phi$ can be taken to be real, since the radial component of the Maxwell equations implies the phase of $\Phi$ is constant. Note that the last two terms in \eqref{EFFFFMASSS} correspond to the effective mass term of $\Phi$ with the effective mass $m_{\text{eff}}^2:=q^2 A^2 +B(\phi)$.

Plugging the following scaling ansatz \eqref{SAS} into \eqref{EFFFFMASSS}, we can study the scaling property of $\Phi$ with the IR geometry \eqref{ir3charge}
\begin{align} \label{SAS}
\begin{split}
\Phi = \Phi_{0} \, \rho^{\Delta_{\Phi}} \,, \qquad B(\phi) := B_{\text{IR}}(\phi) = B_{0} \, \rho^{\Delta_{B}} \,,
\end{split}
\end{align}
and one can find that the kinetic term and the effective mass term behave as follows
\begin{align} \label{}
\begin{split}
(\partial \Phi)^2 \,\sim\, \rho^{2\Delta_{\Phi}+1} \,, \qquad q^2 \,A^2\, \Phi^2 \,\sim\, \rho^{2\Delta_{\Phi}-1} \,, \qquad B_\text{IR}(\phi)\,\Phi^2 \,\sim\, \rho^{2\Delta_{\Phi}+\Delta_{B}} \,.
\end{split}
\end{align}

In general, one can notice that the gauge field contribution to the effective mass, $q^2 A^2 \Phi^2$ term, scales differently with the kinetic term, $(\partial \Phi)^2$. However, the contribution from the coupling $B(\phi)$, $B_\text{IR}(\phi)\,\Phi^2$, can scale in the same way as the kinetic term if
\begin{align} \label{sCSSC}
\begin{split}
\text{Scaling case}: \qquad \Delta_{B} = 1 \qquad \Leftrightarrow \qquad B_{\text{IR}}(\phi) \,=\, B_0 \, \rho \,\sim\, e^{\frac{\phi}{\sqrt{3}}} \,,
\end{split}
\end{align}
where a dilaton solution in \eqref{ir3charge} is used in the proportionality. This \eqref{sCSSC} is called ``the scaling case"~\cite{Cremonini:2016bqw}.

One can generalize \eqref{sCSSC} for the generic scaling IR behavior of $B(\phi)$ with the one parameter $\tau$ as
\begin{align} \label{BIR}
\begin{split}
B_{\text{IR}}(\phi) = B_{0} \, e^{\tau  \phi} \,,
\end{split}
\end{align}
where $\tau = 1/\sqrt{3}$ corresponds to \eqref{sCSSC}.

\paragraph{The UV-completed coupling $B(\phi)$: }
In principle, there would be many possibilities to choose the UV-completed coupling, $B(\phi)$, satisfying \eqref{BIR} in IR. In this paper, for concreteness in our discussion and numerics, we choose one minimal way studied in \cite{Cremonini:2016bqw}:
\begin{align} \label{COUPLING}
B(\phi) = M^{2} \, \cosh \left( \tau \phi \right) \,,
\end{align}
where it has two parameters ($M, \tau$). Note that the coupling \eqref{COUPLING} is the same form of the dilaton potential in \eqref{action1}: $\cosh(\phi/\sqrt{3})$.
In UV region ($r\rightarrow\infty$) (or $\phi\rightarrow 0$), the coupling \eqref{COUPLING} is expanded as 
\begin{equation}\label{ASYMPB}
 B_{\text{UV}}(\phi) \,\sim\, M^2 \left(1\,+\,\frac{{\tau}^2}{2}\phi^2\,+\,\cdots\right) \,,
\end{equation}
and in IR region ($r\rightarrow0$) (or $\phi\rightarrow \infty$) we have
\begin{equation}
 B_{\text{IR}}(\phi) \,\sim\, \frac{M^2}{2} \, e^{\tau \phi} \,,
\end{equation}
where $B_{0}$ in \eqref{BIR} is $M^2/2$. 
For concreteness in our numerics, we fix $M^{2}=-2$ and $q=6$ in this paper. 
We also make some further comments on $M^2$ at the end of this section. 
Note that we choose the same value for the charge of $\Phi$, $q$, which was used in previous studies of Homes' law~\cite{Erdmenger:2015qqa,Kim:2016jjk,Kim:2016hzi} for an easy comparison.

\subsubsection{Superconducting instability with $T_{c}$}

Now we investigate the critical temperature $T_{c}$ with \eqref{COUPLING}, which might be important not only for Homes' law, but also for the study of high $T_c$ superconductors.
We will determine $T_{c}$ by solving the complex scalar field equation of motion \eqref{CSFEOM}:
\begin{equation}\label{mode0}
\Phi''(r)+\left(\frac{f'}{f}+\frac{h'}{h}\right)\Phi'(r)-\frac{1}{f}\left(B(\phi)-\frac{q^2 A_t^2}{f}\right)\Phi(r)=0 \, ,
\end{equation}
where we can use $f(r)$, $h(r)$, and $A_{t}(r)$ from \eqref{3charge} in the absence of the back-reaction.

In order to solve the equation of motion \eqref{mode0}, we impose two boundary conditions. The first condition is from the horizon, $r_{h}$, with the regularity condition in which $\Phi'(r_{h})$ will be determined by $\Phi(r_{h})$. The second boundary condition comes from the AdS boundary, $r\rightarrow\infty$. $\Phi$ behaves near the AdS boundary as
\begin{equation}\label{M2CONFORMAL}
\Phi(r) \,=\, \frac{\Phi^{(-)}}{r^{\Delta_-}} \,+\, \frac{\Phi^{(+)}}{r^{\Delta_+}} \,+\, \cdots,\quad \Delta_{\pm}=\frac{3\pm\sqrt{4 M^2 + 9}}{2} \, .
\end{equation}
By the holographic dictionary, the fast falloff of the field $\Phi^{(-)}$ is interpreted as the source and the slow falloff $\Phi^{(+)}$ corresponds to the condensate.
As a boundary condition for superconductors, we set the source term, $\Phi^{(-)}$, to be zero to describe a spontaneous symmetry breaking. Thus, when $\Phi^{(+)}$ is finite the state will be a superconducting phase, while if $\Phi^{(+)}=0$ (or $\Phi = 0$) the state corresponds to a normal phase.

\paragraph{The critical temperature $T_{c}$ vs $k/\mu$:}
Solving equation of motion \eqref{mode0} with the boundary conditions above, one can find $T_c$ at which the condensate $\Phi^{(+)}$ starts to be finite. 

\begin{figure}[]
\centering
     {\includegraphics[width=7.8cm]{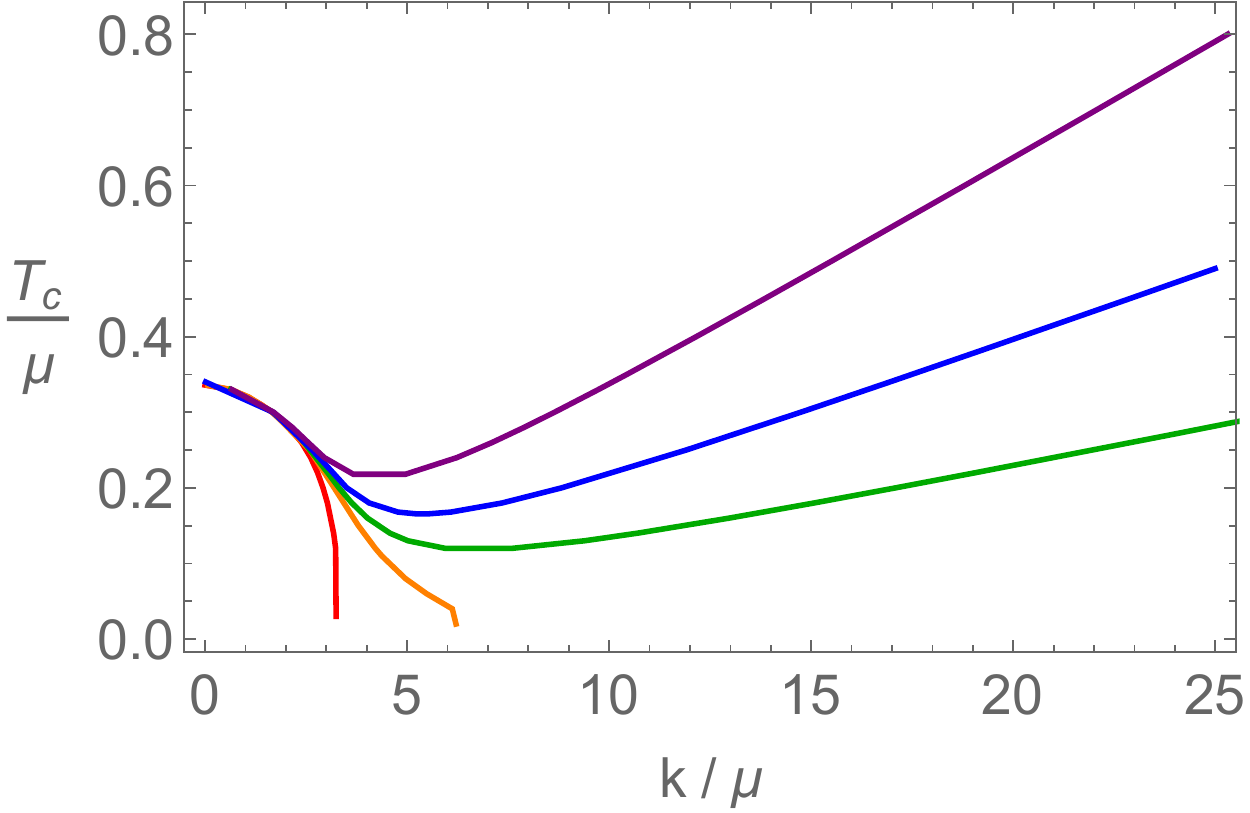}}
\caption{$T_{c}$ vs $k/\mu$ with $\tau  \in \text{[}0, \,\text{}\frac{1}{\sqrt{3}}\frac{12}{10}]$ (red-purple).} \label{TCFIG}
\end{figure}

In Fig. \ref{TCFIG}, we display the plot for $T_c$ in terms of $k/\mu$ with various $\tau$.
In $k/\mu\ll1$ region, one cannot find the $\tau$ effect on $T_c$, i.e., $T_c$ is independent on $\tau$ in the coherent regime  ($k/\mu\rightarrow0$). This would be consistent with the result at $k/\mu=0$ in \cite{Cremonini:2016bqw}.\footnote{Depending on the parameter regime in ($q$, $M^2$), there would be a minimal charge $q$ below which the superconducting instability with $T_c$ does not exist~\cite{Cremonini:2016bqw}. Similar parameter regime may also appear in the presence of momentum relaxation, we leave it as future work.}

However, as $k/\mu$ is increased, we find two main features related to $\tau$. 
First, there would be a critical $\tau$, $\tau_{c}$, to study superconducting phases at $k/\mu\gg1$ limit. 
For instance, if $\tau=\frac{1}{\sqrt{3}}\frac{8}{10}$ (orange)  in Fig. \ref{TCFIG}, superconducting phases cannot be obtained in $k/\mu>6$ regime.
Therfore, in order to study superconductors in the strong momentum relaxation limit ($k/\mu\rightarrow\infty$), we need to consider $\tau>\tau_{c}$. In this paper, we take $\tau_{c}=\frac{1}{\sqrt{3}}\frac{8}{10}$ for simplicity: in Fig. \ref{TCFIG}, one may try to find a more exact value for $\tau_{c}$ between the orange one ($\tau_{c}=\frac{1}{\sqrt{3}}\frac{8}{10}$) and the green one ($\tau_{c}=\frac{1}{\sqrt{3}}\frac{9}{10}$). 

Second, at given $k/\mu$, $\tau$ enhances $T_c$ (e.g., from green to purple). This indicates that the superconducting instability can be triggered more easily at higher coupling $\tau$. Thus, a larger $\tau$ might be useful to investigate the superconducting phase at higher temperature, i.e., high $T_c$ superconductors.

\paragraph{High $T_{c}$ superconductor and linear-$T$ resistivity:}
In summary, we find that the coupling $\tau$, $\tau>\tau_{c}$, would be important not only for superconducting phases at strong momentum relaxation region, but also for high $T_c$ superconductors.

Based on this result, we may argue that, in order to describe high $T_{c}$ superconductors having linear-$T$ resistivity \eqref{sigmabigb} near $T_c$ (i.e., the region where the normal phase still can be useful), we may need the following conditions:
\begin{align}\label{CD22}
\text{i)} \,\, \tau>\tau_{c};   \qquad\qquad  \text{ii)} \,\,  k/\mu\gg1 \,.
\end{align}
This would imply that the trivial coupling term $B(\phi)=M^2$ ($\tau=0$ case) used in most of the previous literature may not capture a complete feature of the superconducting phases at strong momentum relaxation limit. 
In the following section \ref{sec3}, we will study if the condition \eqref{CD22} can also be related to Homes' law.

\paragraph{Instability condition for $M^2$:}
We finish this section with the instability condition for $M^2$ with the complex scalar field equation of motion. Here we consider the scaling case ($\tau=\frac{1}{\sqrt{3}}$) because one can obtain a simple analytic instability condition with it.\footnote{The same thing happens even at $k/\mu=0$ in \cite{Cremonini:2016bqw}. We cannot find the analog of the generalized BF bound unless $\tau=\frac{1}{\sqrt{3}}$. So, identifying the analytic expression for the onset of the phase transition for non-scaling case is still challenging.} 

At $\tau=\frac{1}{\sqrt{3}}$, the complex scalar field equation of motion in the IR geometry \eqref{ir3charge} can be expressed as
\begin{equation}
\Phi''(\rho) - \frac{1}{\rho}\Phi'(\rho) +  \frac{4Q^2 \left( M^2(k^2-6Q^2)\rho^2 - 2q^2(k^2-2Q^2) \right)}{(k^2-6Q^2)^2 \rho^4}  \, \Phi(\rho)=0 \, ,
\end{equation}
and this equation can be solved analytically by the combination of Bessel functions $J_{\nu}$:
\begin{equation}\label{scalepsi}
\Phi(\rho) = c_{1} \,\rho\, \Gamma (1-\nu) J_{-\nu}\left(\sqrt{\frac{8q^2 Q^2 (k^2-2Q^2)}{(k^2-6Q^2)^2}}\frac{1}{\rho}\right) \,+\, c_{2}\,\rho\, \Gamma (1+\nu) J_{\nu}\left(\sqrt{\frac{8q^2 Q^2 (k^2-2Q^2)}{(k^2-6Q^2)^2}}\frac{1}{\rho}\right) \, ,
\end{equation}
where the index $\nu$ is
\begin{align}\label{}
\begin{split}
 \nu=\sqrt{\frac{k^2 - (6+4M  ^2)Q^2}{k^2-6Q^2}}\,.
\end{split}
\end{align}
Then, the instability appears when the index $\nu$ of Bessel function becomes imaginary. 
Note that, unlike the case of $k=0$ in \cite{Cremonini:2016bqw}, there are two ways to make the  imaginary $\nu$ depending on the sign of the numerator (or denominator) in $\nu$: i) the positive numerator with the negative denominator; ii) the negative numerator with the positive denominator. Each case produces the following instability condition for $M^2$
\begin{align}\label{INSTAM}
\begin{split}
 M^2 < \frac{k^2 - 6Q^2}{4Q^2} < 0  \qquad\text{or}\qquad M^2 > \frac{k^2 - 6Q^2}{4Q^2}  > 0 \,.
\end{split}
\end{align}
Note that only the first condition in \eqref{INSTAM} is consistent at $k=0$. Thus, from the perspective of continuity encompassing the $k=0$ case, the first condition in \eqref{INSTAM} might be the proper instability condition.

%
\section{Homes' law}\label{sec3}
In this section, considering the coupling \eqref{COUPLING}, we study Homes' law with the fully back-reacted geometry.

\subsection{Setup for numerics}
We consider the following ansatz to obtain the fully back-reacted background solutions 
\begin{equation} \label{ANSATZNUM}
\begin{split}
&\dd s^2 = \frac{1}{\tilde{z}^2} \left[-(1-\tilde{z})U(\tilde{z}) \dd \tilde{t}^2+\frac{\dd \tilde{z}^2}{(1-\tilde{z})U(\tilde{z})}+V(\tilde{z})\dd \tilde{x}^2 +V(\tilde{z}) \dd \tilde{y}^2 \right] \,, \\
&A=(1-\tilde{z})a(\tilde{z}) \dd \tilde{t} \,, \quad  \phi=\frac{\sqrt{3}}{2} \log[1+\varphi(\tilde{z})] \,, \quad \Phi=\tilde{z}^{\Delta_{-}}\, \eta(\tilde{z}), \\
&\psi_{1}=\tilde{k} \, \tilde{x} \,, \quad \psi_{2}=\tilde{k} \, \tilde{y} \,,
\end{split}
\end{equation}
where 
\begin{equation} \label{tildes}
\tilde{z} :=  \frac{z}{z_{h}} \,, \qquad \tilde{t} :=\frac{t}{ z_{h}} \,, \qquad \tilde{x} :=\frac{x}{z_{h}} \,, \qquad \tilde{y} :=\frac{y}{z_{h}} \,, \qquad  \tb :=  k \, z_{h} \,.  
\end{equation}
Here $U ,V, a, \varphi$ and $\eta$ are functions of the holographic direction $\tilde{z}$. In this coordinate, the AdS boundary is located at $\tilde{z}=0$ and the horizon is at $\tilde{z}=1$. Note that the coordinate \eqref{ANSATZNUM} is related to \eqref{3charge} with $z=1/r$ and the form of ansatz \eqref{ANSATZNUM} is chosen for the convenience of numerical analysis for superconducting phase.

With the ansatz \eqref{ANSATZNUM}, one can identify the Hawking temperature $T$ and the chemical potential $\mu$ as  
\begin{align}
T= \frac{g_{tt}'(\tilde{z})}{4\pi \sqrt{g_{tt} g_{\tilde{z}\tilde{z}}}}\Bigr|_{\tilde{z}=1} = \frac{U(1)}{4\pi} \,, \quad \mu= A_t(0) = a(0) \,,
\end{align}
and the condensate $\Phi^{(+)}$ in \eqref{M2CONFORMAL} can be read off from $\eta$  in \eqref{ANSATZNUM} near the AdS boundary as 
\begin{align} \label{}
\begin{split}
 \Phi = \Phi^{(-)} \, \tilde{z}^{\Delta_{-}} \,+\, \Phi^{(+)} \, \tilde{z}^{\Delta_{+}} \,+\, \dots  \quad \Leftrightarrow \quad \eta = \Phi^{(-)}  \,+\, \Phi^{(+)} \, \tilde{z}^{\Delta_{+}-\Delta_{-}} \,+\, \dots  \,,
\end{split}
\end{align}
where $\Delta_{\pm}$ is defined in \eqref{M2CONFORMAL}.\footnote{For $M^2=-2$, $(\Delta_{-}, \Delta_{+})=(1,2)$.}

\begin{figure}[]
 \centering
     {\includegraphics[width=7.8cm]{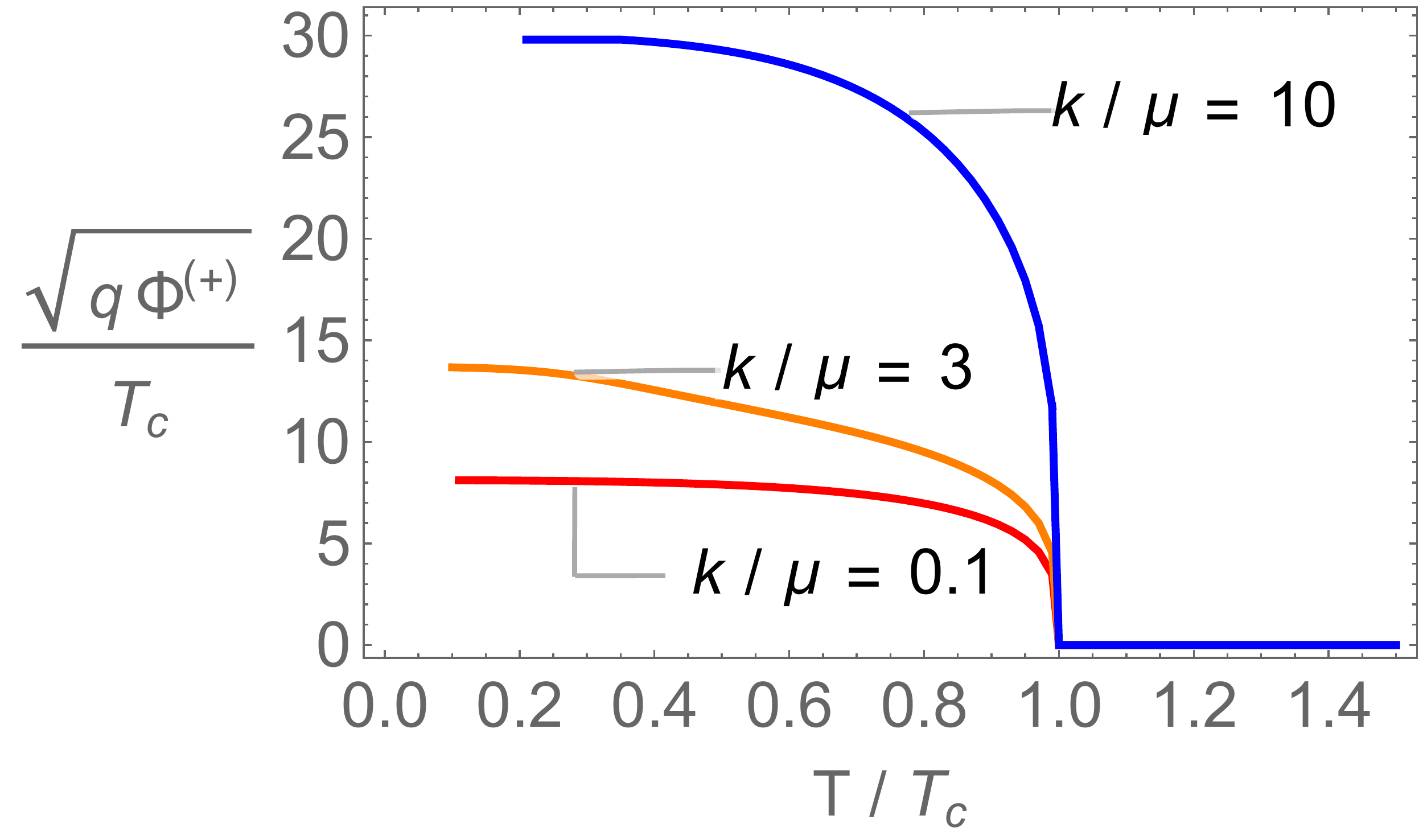} \label{}}
          \caption{Condensation vs temperature at $\tau=1/\sqrt{3}$ where $T_c$ is the critical temperature. } \label{CONPLOT}
\end{figure}

In order for the superconducting phases, we set the source $\Phi^{(-)} = 0$, i.e., $\eta(0) = 0$ and need to find the state with the finite condensate $\Phi^{(+)} \neq 0$.
One can find such a state by solving the equations \eqref{MATTEREOM}-\eqref{EiensteinEq} numerically with the ansatz \eqref{ANSATZNUM}. The typical condensate is plotted in Fig. \ref{CONPLOT}: the condensate tends to be enhanced with increasing $k/\mu$ and this seems to be a generic feature of holographic superconductors in the presence of the momentum relaxation.

In what follows, in section \ref{sdoc32}, we focus on the scaling case $\tau=1/\sqrt{3} > \tau_{c}$ and study the electric optical conductivity $\sigma(\omega)$ and the superfluid density $\rho_{s}$. Then, using the results in section \ref{sdoc32}, we investigate Homes' law in section \ref{hlsme}. In section \ref{taudhw}, we discuss the $\tau$ effect on Homes' law.

\subsection{Electric conductivity and superfluid density}\label{sdoc32}

Let us study the electric optical conductivity of the holographic model \eqref{action1}. From here on, we use the scaled variables \eqref{tildes} without tilde for simplicity. 

\paragraph{Holographic electric optical conductivity:} In order to compute the electric optical conductivity, we need to consider the following fluctuations:
\begin{align} \label{FLUCVaria}
\begin{split}
\delta g_{tx} = h_{tx}(z) \, e^{-i \omega t} \,, \qquad \delta A_{x} = a_{x}(z) \, e^{-i \omega t} \,, \qquad \delta \psi_{x} = \xi_{x}(z) \, e^{-i \omega t} \,,
\end{split}
\end{align}
where the fluctuations behave near the AdS boundary as
\begin{align} \label{FLUCBOUND}
\begin{split}
h_{tx}(z) &\,=\, \frac{h_{tx}^{(S)}}{z^2} \,+\, h_{tx}^{(R)} \,+\, \dots   \,, \\
a_{x}(z) &\,=\,  a_{x}^{(S)}   \,\,\,+\,\, a_{x}^{(R)}\, z \,+\, \dots \,,  \\ 
\xi_{x}(z) &\,=\, \xi_{x}^{(S)} \,\,\,+\,\, \xi_{x}^{(R)}\, z \,+\, \dots \,, 
\end{split}
\end{align}
here the leading coefficients ($h_{tx}^{(S)}, a_{x}^{(S)}, \xi_{x}^{(S)}$) correspond to the sources, and the subleading terms ($h_{tx}^{(R)}, a_{x}^{(R)}, \xi_{x}^{(R)}$) would be interpreted as the response by the holographic dictionary.

The electric optical conductivity can be obtained by the Kubo formula in terms of the boundary coefficients in \eqref{FLUCBOUND}:
\begin{align} \label{KBF}
\begin{split}
\sigma(\omega) = \frac{1}{i\omega} G_{j^{x}j^{x}}^{R}(\omega) = \frac{a_{x}^{(R)}}{i\omega \, a_{x}^{(S)}} \,,
\end{split}
\end{align}
where $G_{j^{x}j^{x}}^{R}$ is the current-current retarded Green's function. 
The second equality in \eqref{KBF} holds when $a_{x}^{(S)}$ is the only non-zero source.

In order to make the source-vanishing boundary condition except $a_{x}^{(S)}$, one may use the diffeomorphisms and gauge-transformations~\cite{Kim:2015sma,Donos:2013eha}.
With a constant residual gauge parameter $\zeta$ fixing $\delta g_{rx}=0$~\cite{Kim:2015sma}, it can be shown that the fluctuations in \eqref{FLUCVaria} except $\delta A_x$ can be expanded near the AdS boundary as\footnote{For more detailed analysis and discussion about the diffeomorphism and gauge-transformations, see \cite{Kim:2016hzi,Kim:2015sma,Donos:2013eha}.}
\begin{align} \label{OTHERSOURCE}
\begin{split}
z^2 \, \delta {g_{tx}} \,\sim\, (h_{tx}^{(S)} - i \omega \zeta) e^{-i\omega t} \,=\,0 \,, \qquad \delta \psi_{x} \,\sim\, (\xi_{x}^{(S)} + k \zeta) e^{-i\omega t} \,=\,0 \,,
\end{split}
\end{align}
where the equalities correspond to the source-vanishing boundary condition except $a_{x}^{(S)}$.
Plugging one of the equations in \eqref{OTHERSOURCE} into the other, \eqref{OTHERSOURCE} produces a single condition
\begin{align} \label{DFF}
\begin{split}
\xi_{x}^{(S)} - \frac{i k}{\omega} h_{tx}^{(S)}  = 0\,.
\end{split}
\end{align}
Therefore, one can use \eqref{KBF} to study the electric optical conductivity by solving the equation of motion of the fluctuations with the boundary condition \eqref{DFF}.

\begin{figure}[]
 \centering
     {\includegraphics[width=4.831cm]{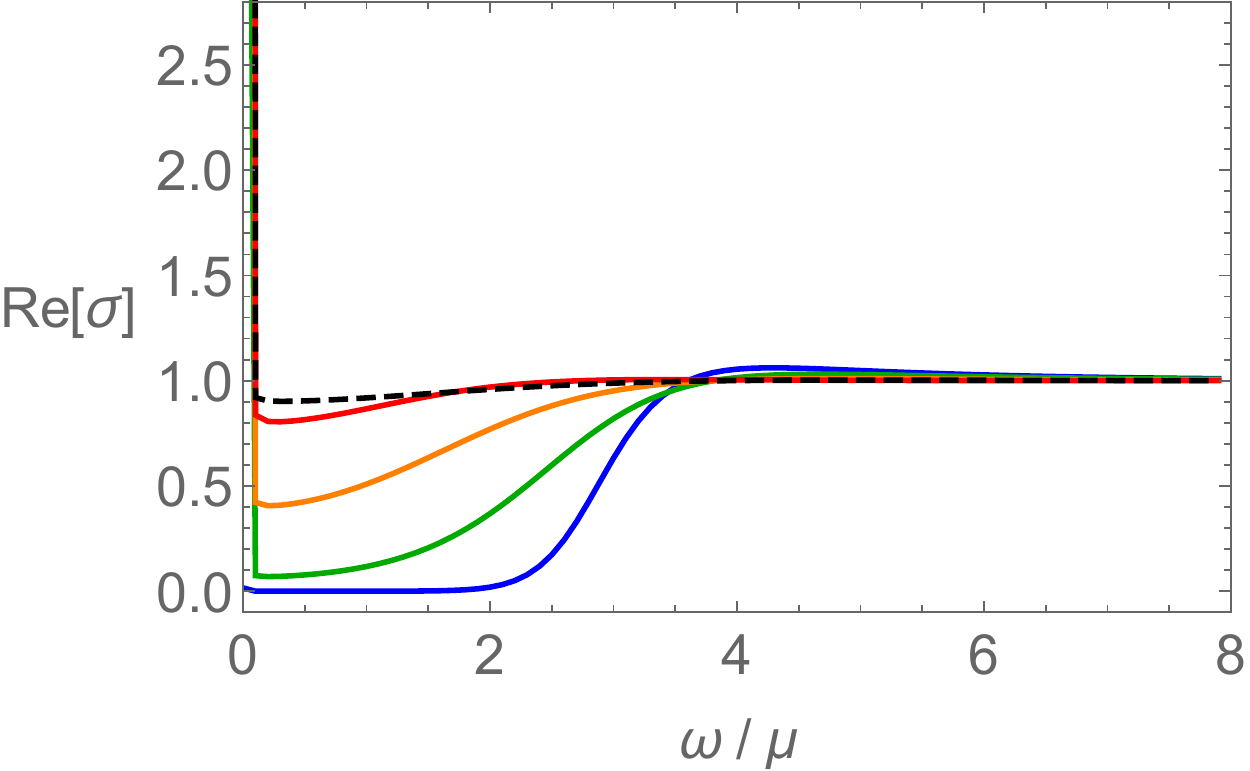} \label{}}
     {\includegraphics[width=4.831cm]{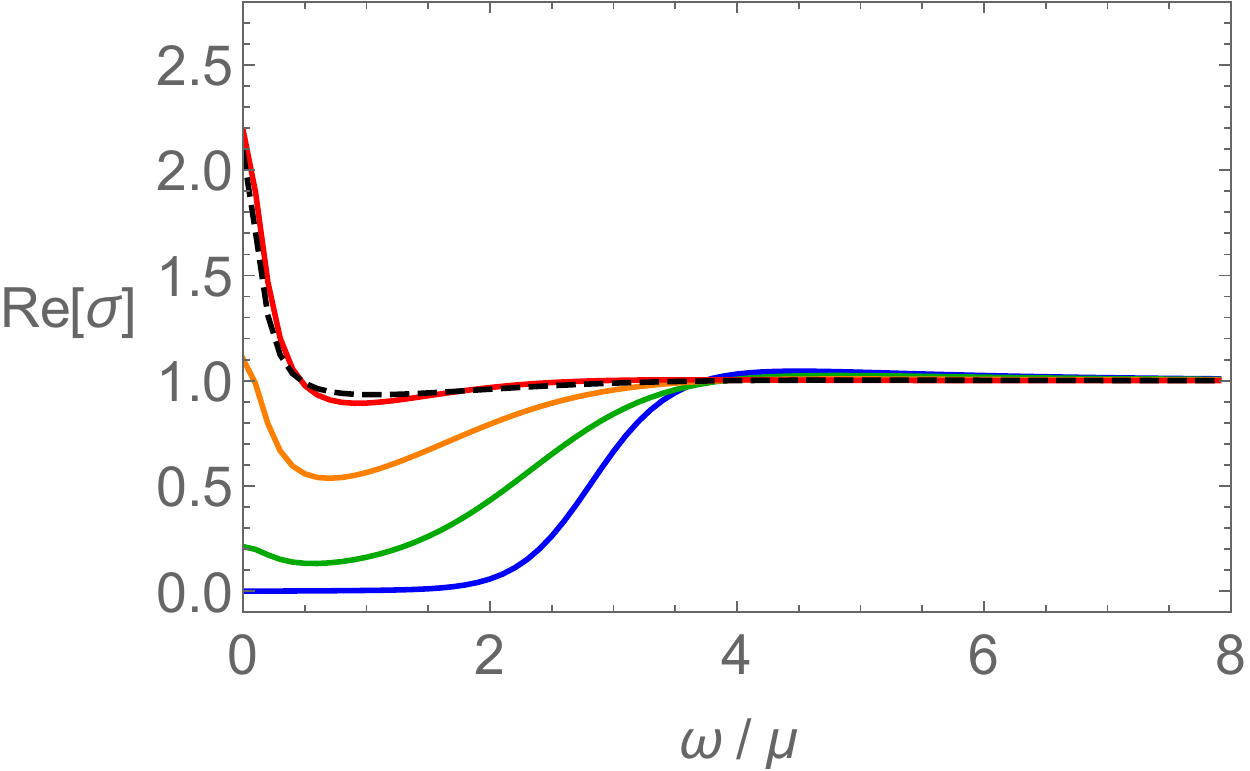} \label{ACAC2}} 
     {\includegraphics[width=4.831cm]{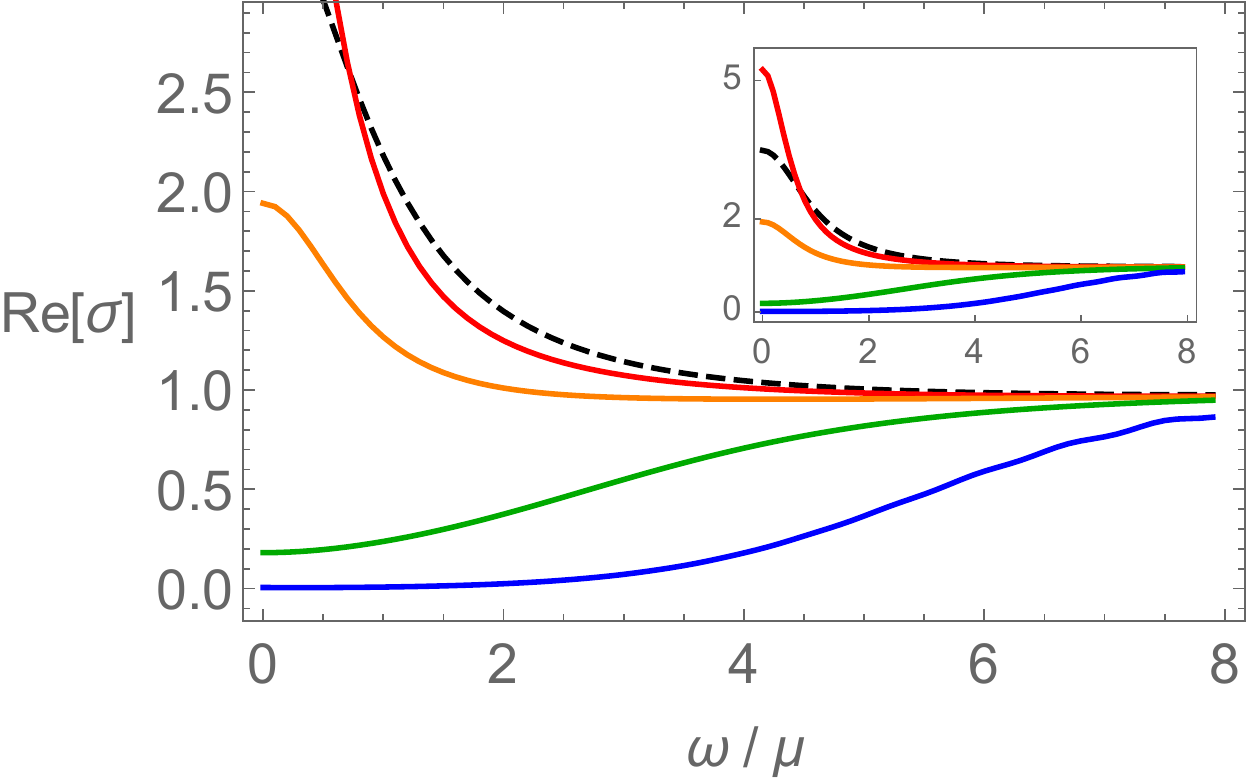} \label{ACAC3}}
     
     \subfigure[$ k/\mu=0.1, \ \  T/T_c = 1.5, 1, 0.93,$ $0.75, 0.21$ (dashed, red, orange, green, blue)] 
     {\includegraphics[width=4.831cm]{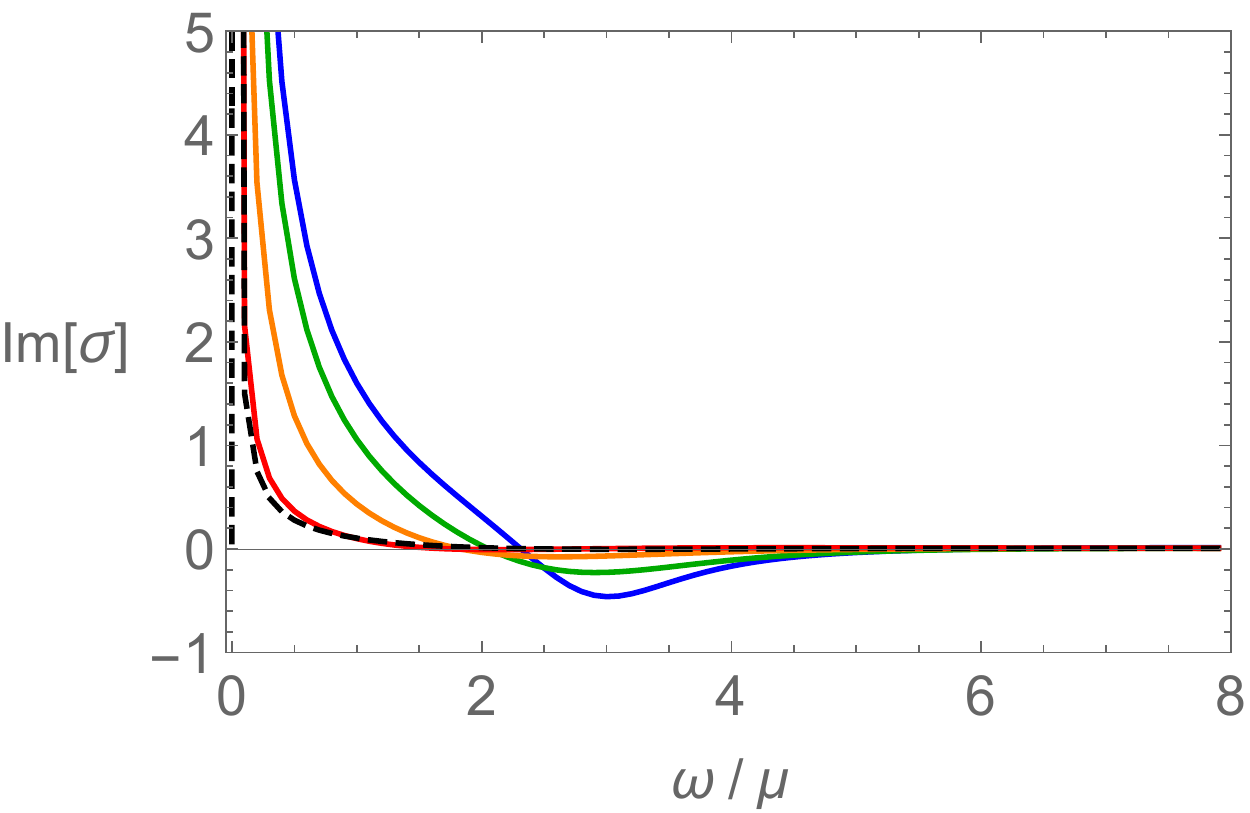} \label{ACACa}}
     \subfigure[$ k/\mu=1,    \ \  T/T_c = 1.5, 1, 0.93,$ $0.75, 0.21$ (dashed, red, orange, green, blue)] 
     {\includegraphics[width=4.831cm]{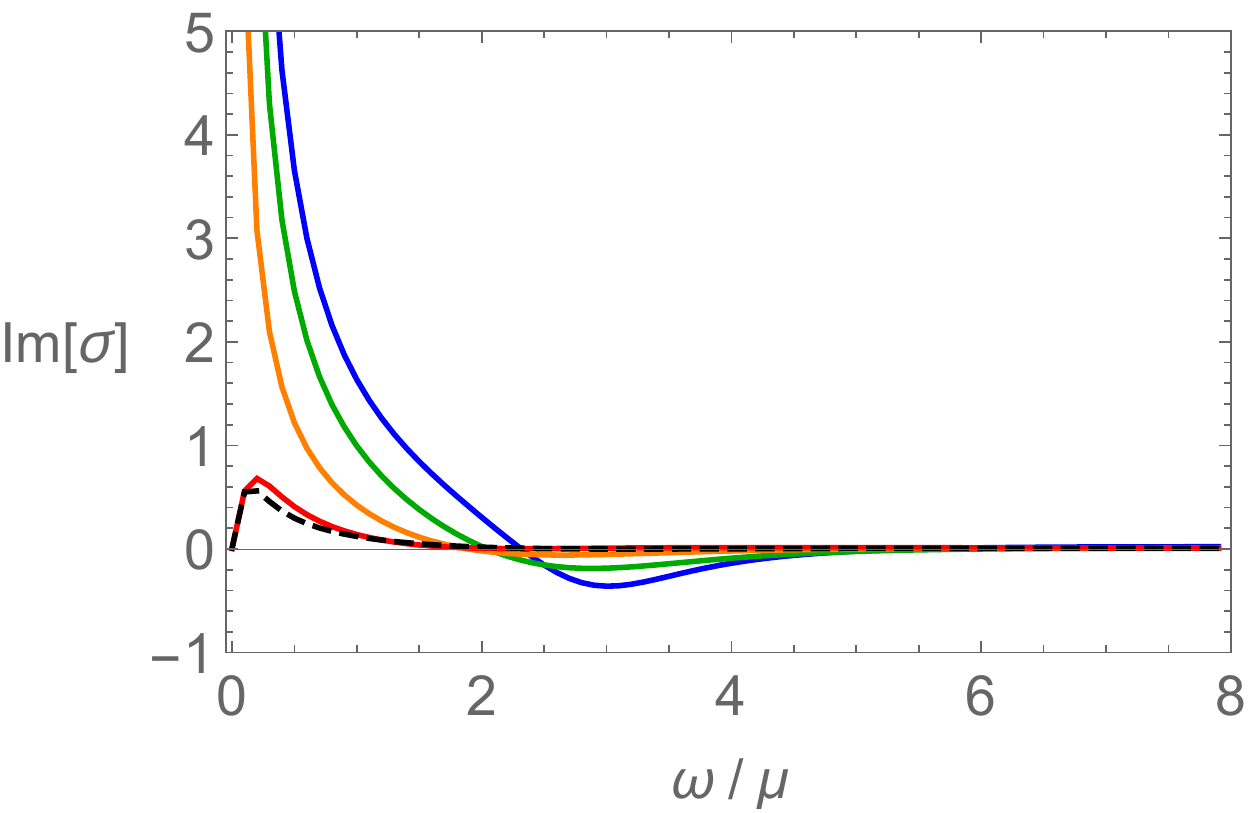} \label{ACACb}} 
     \subfigure[$k/\mu=10,   \ \  T/T_c = 1.5, 1, 0.97,$ $0.87, 0.35$ (dashed, red, orange, green, blue)] 
     {\includegraphics[width=4.831cm]{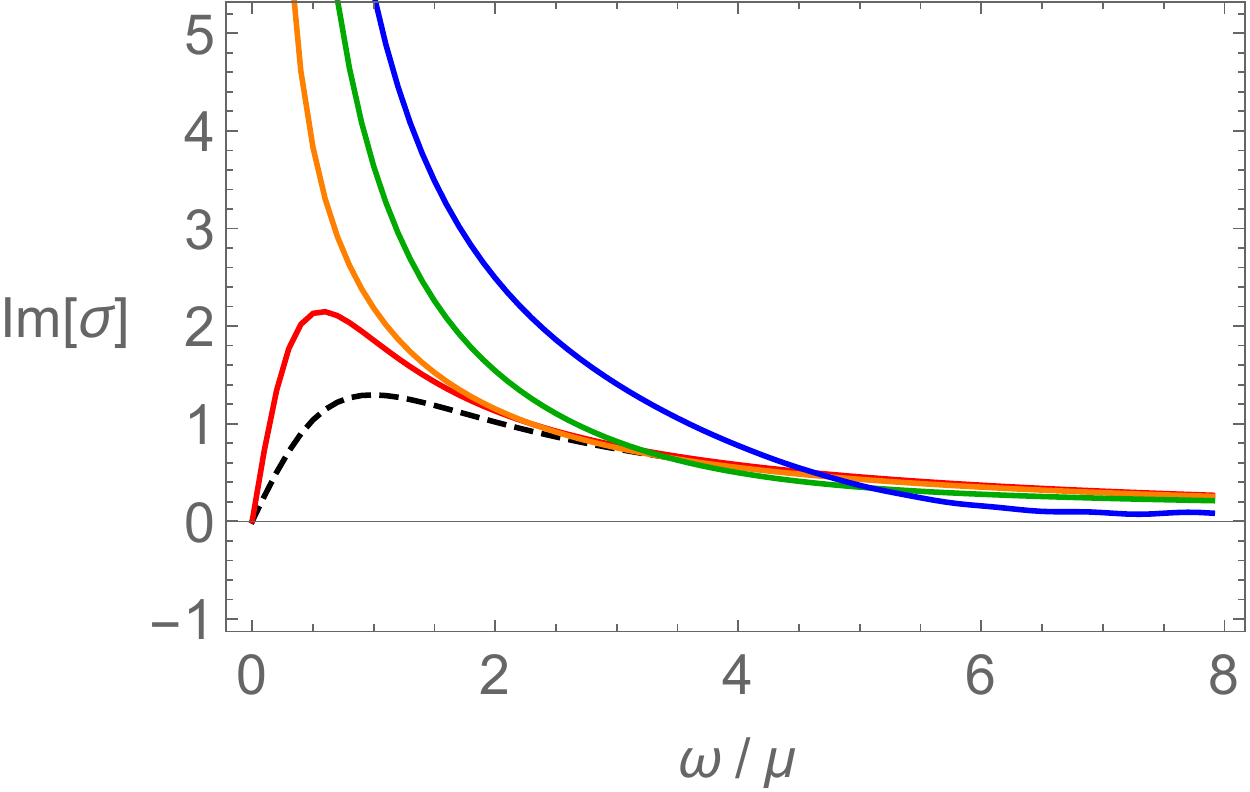} \label{ACACc}}
          \caption{Electric conductivity with various $k/\mu=0.1, 1$ and $10$. The first row displays Re$[\sigma(\omega)]$, while Im$[\sigma(\omega)]$ is plotted in the second row ($\tau=1/\sqrt{3}$).} \label{ACAC}
\end{figure}

\paragraph{Electric optical conductivity in strange metal/superconductor transition:}
Using the method above, we make the plot of the optical conductivity in Fig. \ref{ACAC}.\footnote{We also checked that our numerical code produces the consistent result in \cite{Ling:2013nxa}: the optical conductivity of the Gubser-Rocha model at $k/\mu=0$.}
The color of curves denotes a temperature ratio $T/T_c$: the dashed black is for the normal metal phase ($T>T_{c}$), the red line is for the critical temperature ($T=T_{c}$), and other colors (from orange to blue) correspond to the superconducting phase ($T<T_{c}$).

\begin{figure}[]
 \centering
      \subfigure[$\sigma(\omega)$ vs $\omega/k$. The inset is Im $\sigma$.]
     {\includegraphics[width=7.6cm]{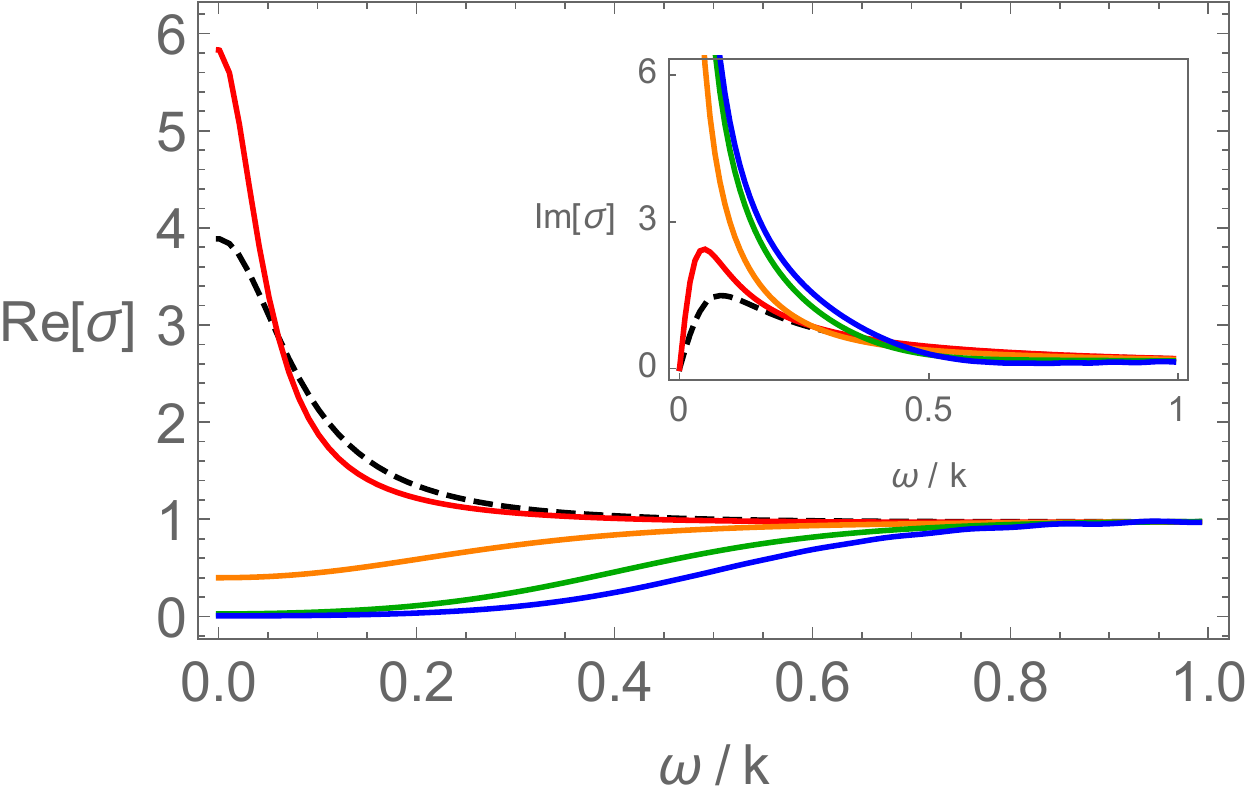}\label{COMPARILAMa}} 
      \subfigure[$\rho$ vs $T/k$.]
     {\includegraphics[width=7.2cm]{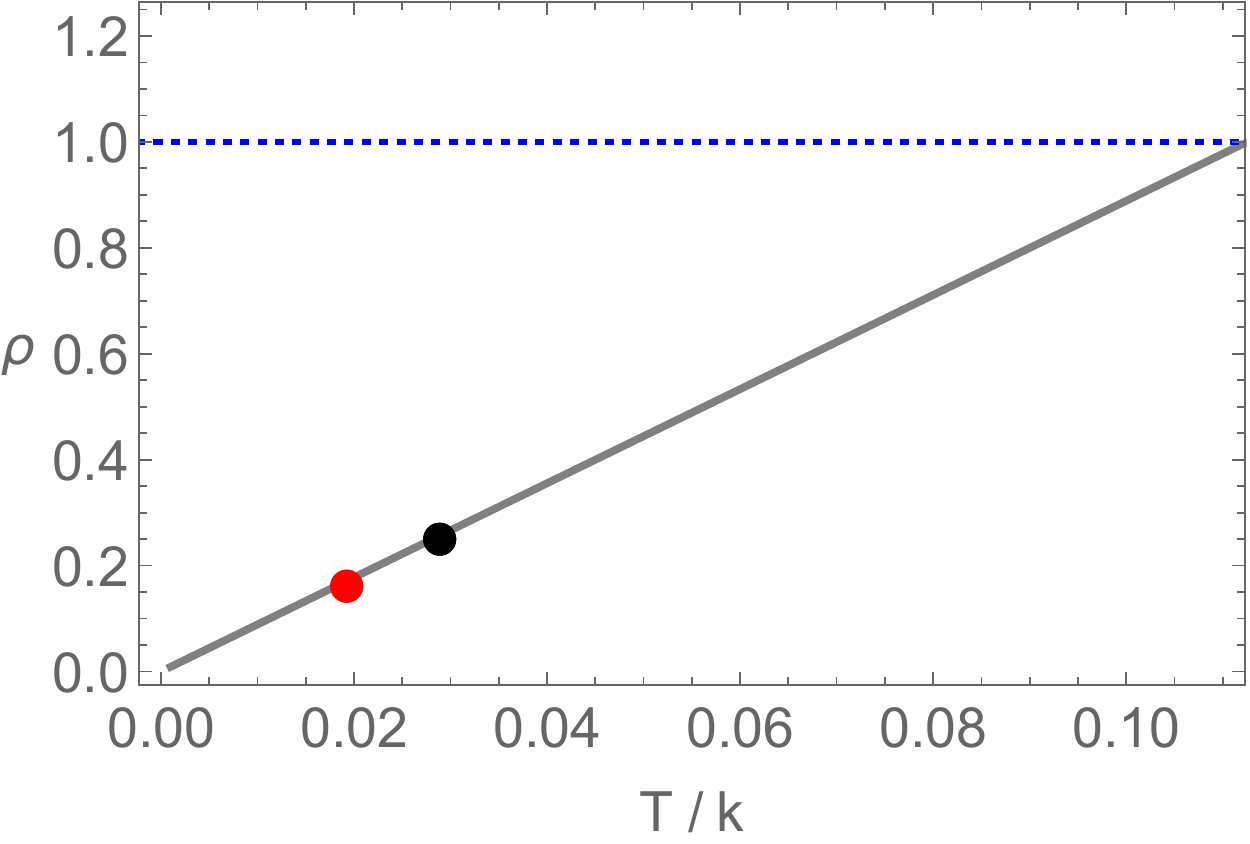} \label{COMPARILAMb}} 
          \caption{The electic conductivity at $\mu/k=0$. \textbf{Left:} $\sigma(\omega)$ of the Gubser-Rocha model at $T/T_c = 1.5, 1, 0.93,$ $0.75, 0.41$ (dashed, red, orange, green, blue). \textbf{Right:} The DC resistivity $\rho$ ($1/\sigma(\omega=0)$) of two holographic models: the Gubser Rocha model (solid gray) \eqref{GUBGUB}, the linear axion model (dotted blue) \eqref{LAMLAM}. The black (red) dot corresponds to the DC limit of the black (red) line in Fig. \ref{COMPARILAMa}: it shows that the numerical result of $\sigma(\omega=0)$ is consistent with the analytic DC result.} \label{COMPARILAM}
\end{figure}

For $T\geq T_{c}$, one can see that the DC conductivity, $\sigma(\omega=0)$, is finite due to the momentum relaxation\footnote{We checked that $\sigma(\omega=0)$ is consistent with the analytic result in \eqref{conduct1}. For instance, see Fig. \ref{COMPARILAMb}.}, while the superconducting phase ($T<T_{c}$) produces $1/\omega$ pole in Im$[\sigma]$ giving the infinite DC conductivity. By the Kramers-Kronig relation, $1/\omega$ pole in Im$[\sigma]$ implies that Re$[\sigma]$ has a delta function at $\omega=0$: this is one of hallmarks of holographic superconductor.

Let us make some further comments on the electric conductivity of our model \eqref{action1}.
In holography, there are two simple gravity models to study the electric conductivity of the normal phase in the presence of the momentum relaxation: i) the linear axion model~\cite{Andrade:2013gsa}\footnote{Recall that the linear axion model is \eqref{action1} without a dilaton field $\phi$.}; ii) the Gubser-Rocha model with the axion field \eqref{action1}.

In \cite{Kim:2015dna}, using the linear axion model, the authors studied the optical conductivity with the phase transition between the normal phase and the superconducting phase.\footnote{Although the linear axion model cannot exhibit the linear-$T$ resistivity, it has been used to describe the metal/superconductor transition.}
Thus, it would be instructive to compare the features in Fig. \ref{ACAC} and the result in \cite{Kim:2015dna}.
To our knowledge, our work is the first holographic study considering the optical conductivity of \eqref{action1}  for the normal phase ($\Phi=0$), i.e., the Gubser-Rocha model with the axion field. 

We found one distinct feature between two holographic models at strong momentum relaxation region: unlike the linear axion model \cite{Kim:2015dna}, at large $k/\mu$, the Drude like peak in the normal phase does not disappear for the Gubser Rocha model (e.g., see dashed black (and red) line in Fig. \ref{ACACc}).
In order to show this feature more clearly at strong momentum relaxation limit ($k/\mu\rightarrow\infty$), we take $\mu/k=0$ and make the plot of the optical conductivity in Fig. \ref{COMPARILAMa}.\footnote{One may directly compare Fig. \ref{COMPARILAMa} with Fig. 4(c) in \cite{Kim:2015dna}.}

The non-vanishing Drude like peak in the strong momentum relaxation limit might be related to the fact that the Gubser-Rocha model produces linear-$T$ resistivity unlike the linear axion model. For instance, at $\mu/k=0$, both holographic models show the DC conductivity as
\begin{align}
&\sigma_{DC} =  \sqrt{1+\tilde{Q}}\left( 1 + \frac{\mu^2}{k^2} \right) \,\sim\, \sqrt{1+\tilde{Q}}  \,\sim\, \frac{k}{2\sqrt{2}\pi T} \,, \,\,\quad \text{(Gubser-Rocha model)}  \label{GUBGUB}\\
&\sigma_{DC} =   1 + \frac{\mu^2}{k^2} \,\sim\, 1  \,,                              \qquad\qquad\qquad\qquad\qquad\qquad\quad\quad\,\,\, \text{(Linear axion model)}  \label{LAMLAM}
\end{align}
where $\tQ$ from the dilaton field plays an important role for the linear-$T$ resistivity.\footnote{\eqref{tQ3} is used in \eqref{GUBGUB} which corresponds to \eqref{sigmabigb}.}

In Fig. \ref{COMPARILAMb}, we display the DC conductivity of two holographic models: the Gubser Rocha model (solid gray) \eqref{GUBGUB}, the linear axion model (dotted blue) \eqref{LAMLAM}.
The black (red) dot in Fig. \ref{COMPARILAMb} corresponds to the DC limit of the black (red) line in Fig. \ref{COMPARILAMa}: it shows that the numerically computed $\sigma(\omega=0)$ is consistent with the analytic DC result.\footnote{Moreover, also note that the linear-$T$ resistivity is robust above $T_c$ (red dot) in Fig. \ref{COMPARILAMb}, which is similar to experiments.}

Note that, at $\omega/k\rightarrow\infty$, $\sigma(\omega)=1$ in both models. However, in the opposite limit, $\omega/k\rightarrow0$, the linear axion model gives the constant value, i.e., $\sigma(\omega)=1$ \eqref{LAMLAM} unlike the Gubser-Rocha model \eqref{GUBGUB}. Thus, for the linear axion model, the optical conductivity of the normal phase would be a constant, $\sigma(\omega)=1$, in all $\omega$ regime without producing a Drude like peak.

\paragraph{Two-fluid model and superfluid density:}
For superconducting phase (roughly $0.5<T/T_{c}<1$) in Fig. \ref{ACAC}, Re$[\sigma]$ also has a remaining finite value at $\omega=0$ in addition to the delta function by the Kramers-Kronig relation.
This residual Drude-like peak may be interpreted by the two-fluid model~\cite{Horowitz:2013jaa} as a contribution from the normal component in the superconducting phase, which has also been observed in other holographic superconductor models such as linear-axion model~\cite{Kim:2015dna}, Q-lattice model~\cite{Ling:2014laa}, and Helical lattice model~\cite{Erdmenger:2015qqa}.

The two-fluid model demonstrates that the low frequency behavior of the optical conductivity can be fitted with the following formula:
\begin{align}\label{Drude1}
\sigma(\omega) \,=\, i\,\frac{\rho_{s}}{\omega} \,+\, \frac{\rho_{n} \,\tau_r}{1-i\,\omega\,\tau_r} \,+\, \rho_{0} \,,
\end{align}
where $\rho_{s}$ and $\rho_{n}$ are defined as the superfluid density and the normal fluid density.\footnote{$\rho_{s}$ and $\rho_{n}$ are supposed to be proportional to the superfluid density and the normal fluid density from the viewpoint of experiments or there could be prefactor $\pi/2$ in each terms. However, for the theoretical study of conductivity, we define the superfluid density and normal fluid density including all the factors.} $\tau_r$ is the relaxation time. $\rho_{0}$ may be related to the pair creation and can be used to fit the numerical data in the presence of  the momentum relaxation~\cite{Kim:2015dna}.

\begin{figure}[]
 \centering
     \subfigure[$k/\mu=0.1$]
     {\includegraphics[width=4.831cm]{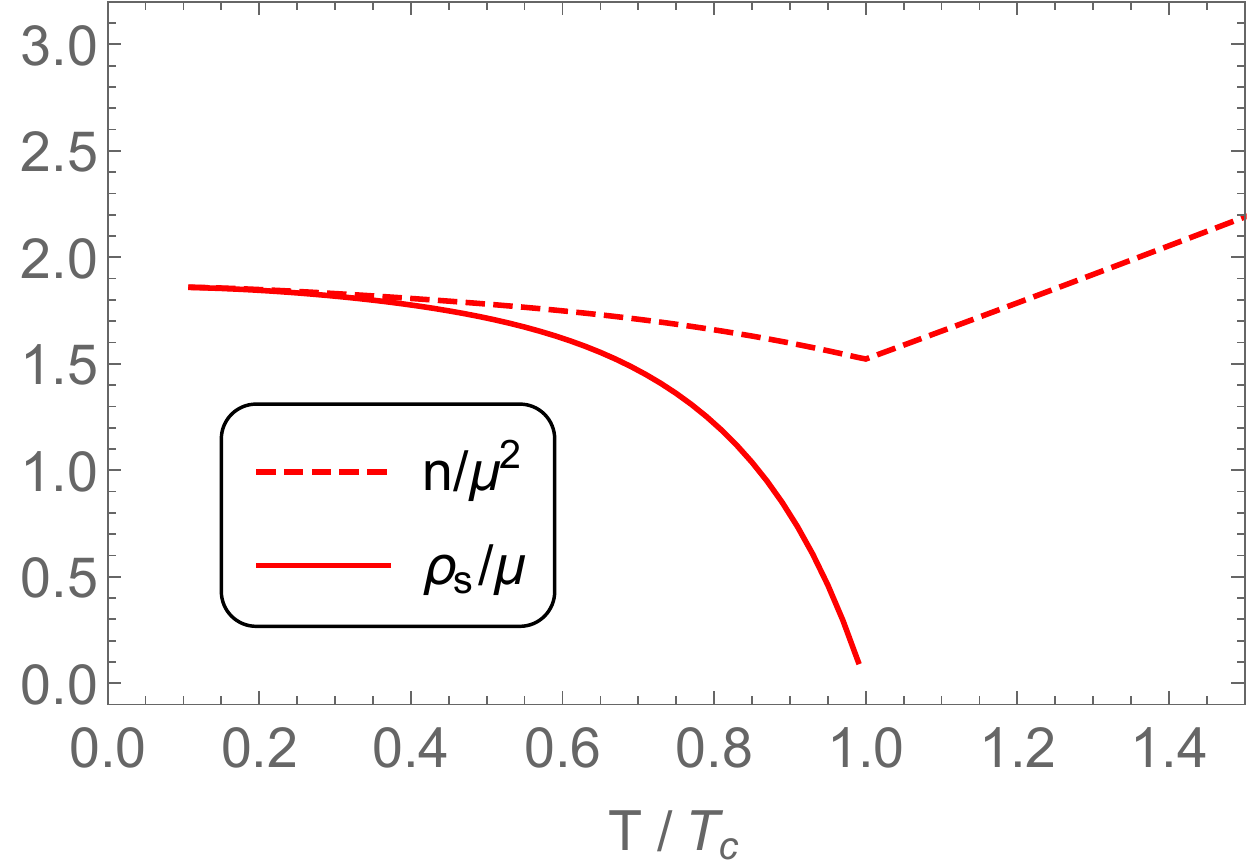} \label{FIG71}}
     \subfigure[$k/\mu=1$]
     {\includegraphics[width=4.831cm]{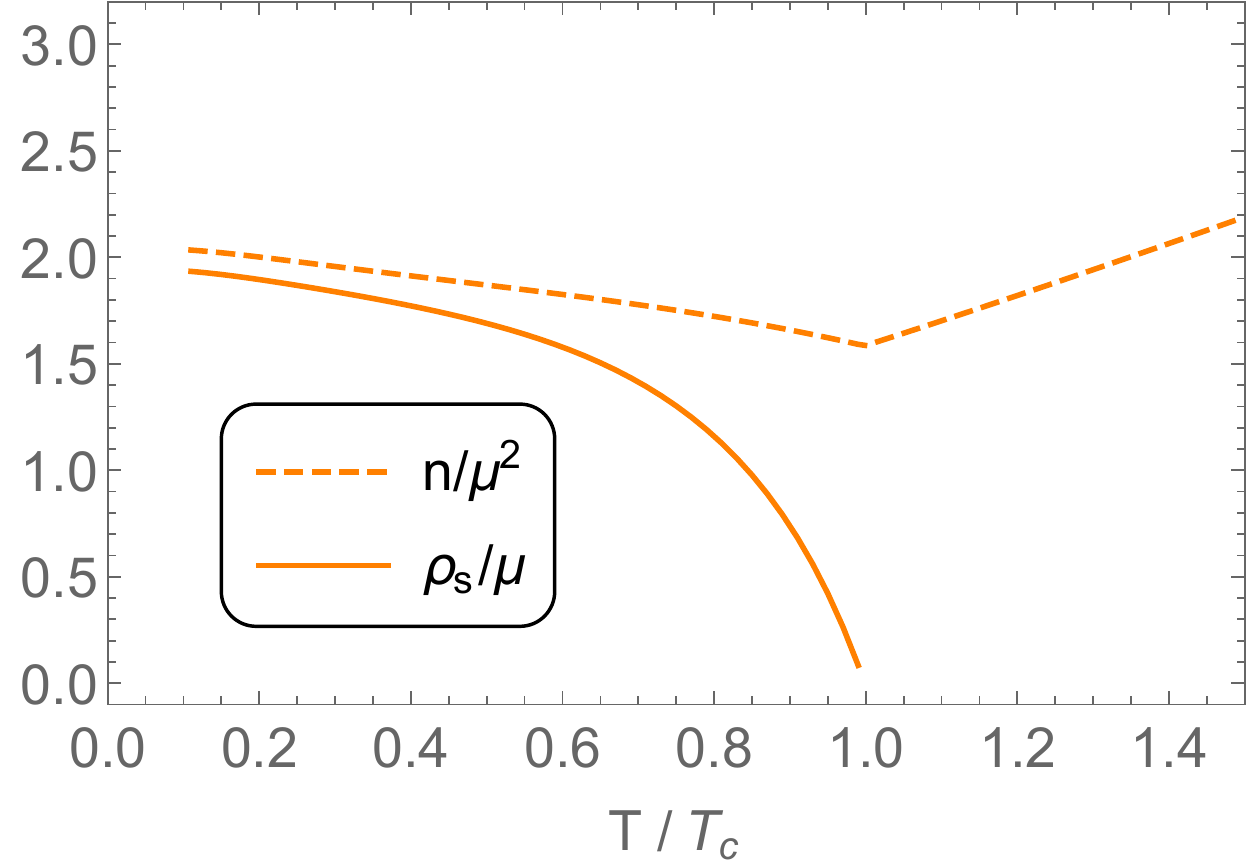} \label{FIG72}} 
     \subfigure[$k/\mu=3$]
     {\includegraphics[width=4.831cm]{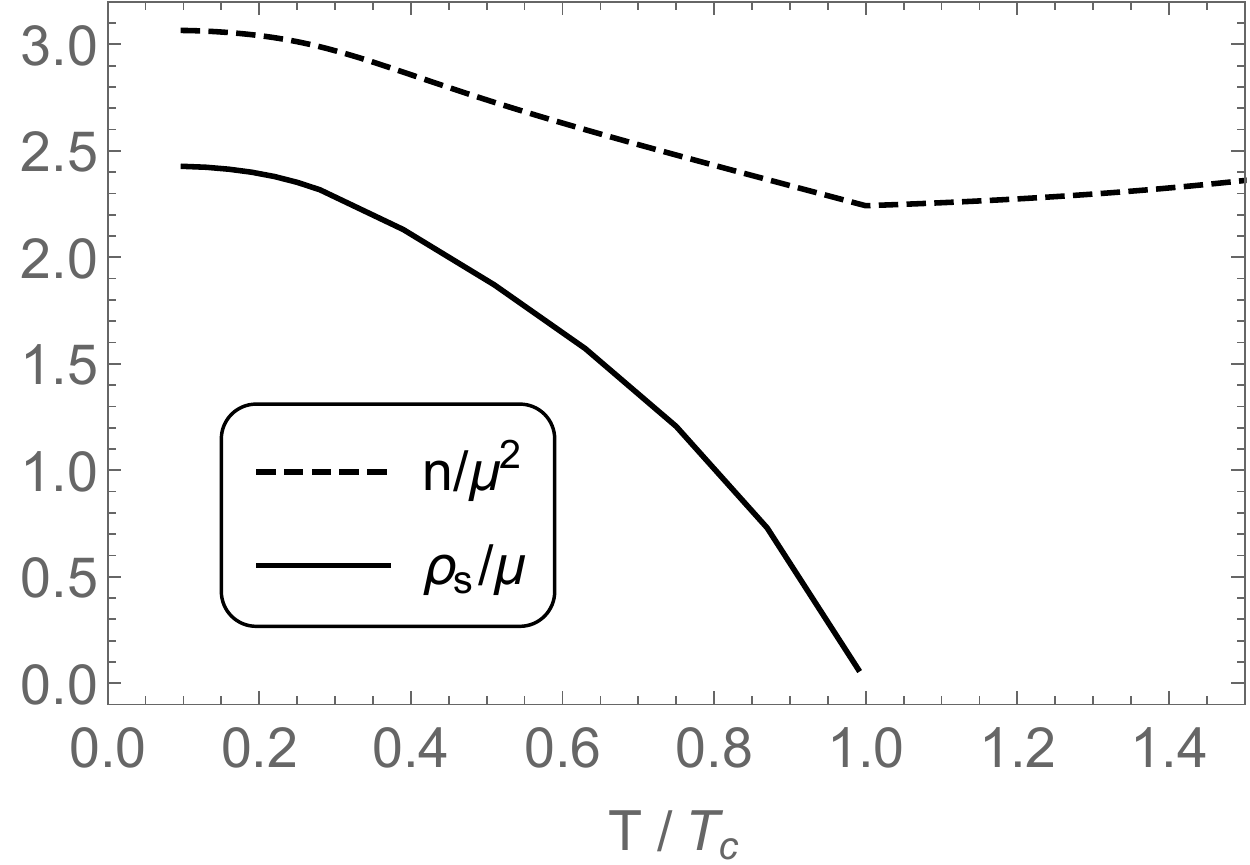} \label{FIG73}}
          \caption{The charge density $n$, superfluid density $\rho_{s}$ vs $T/T_{c}$. The gap between $n$ and $\rho_s$ at $T=0$ opens at finite $k/\mu$ ($\tau=1/\sqrt{3}$).} \label{FIG7}
\end{figure}

The superfluid density $\rho_s$ is our main interest to study Homes' law, which can be read off from the fitting curve \eqref{Drude1}.
One interesting feature of $\rho_s$ in holographic superconductors~\cite{Kim:2016jjk, Erdmenger:2012ik,Kim:2016hzi,Erdmenger:2015qqa} is that there is a finite gap at $T=0$ between $\rho_s$ and the charge density $n$ in the presence of the momentum relaxation.\footnote{The charge density $n$ is defined by a subleading coefficient of $A_t$: $A_t = \mu \,-\, n \, z \,+\, \dots$ near AdS boundary. The normal fluid density would be given by the difference between $n$ and $\rho_s$.}
This also happens in our superconductor model.
See Fig. \ref{FIG7}. Thus, this non-vanishing gap at finite $k/\mu$ seems to be a generic feature of holographic superconductors.

We also find that, as $T$ is lowered, $\rho_{0}$ and $\rho_{n}$ are reduced while $\rho_{s}$ is enhanced.\footnote{At small $T$, $\rho_{0}$ would be vanishing, however $\rho_{n}$ could be finite even at $T=0$~\cite{Horowitz:2013jaa,Zeng:2014uoa,Ling:2014laa}.}
For the relaxation time $\tau_r$, we find it is decreasing as $T$ is lowered. The behavior of $\tau_r$, at low $T$ depends on holographic models: it is increasing in~\cite{Horowitz:2013jaa}, it is decreasing first and then increasing in~\cite{Ling:2014laa}. So, apparently, the more detailed analysis and the unified description for the relaxation time for holographic superconductor is still needed. We leave this subject as future work.

\subsection{Homes' law at strong momentum relaxation}\label{hlsme}

Now let us discuss Homes' law \eqref{HOMEHOME}. Computing three quantities ($\rho_{s}(T=0)$, $T_{c}$, $\sigma_{DC}(T_{c})$) as a function of $k/\mu$, we may check Homes' law at given $\tau$.\footnote{Recall that we have two parameters ($\tau$, $k/\mu$) in our setup.}

Note that $\rho_{s}(T=0)$ can be read off from \eqref{Drude1} in principle.
However, as one can see from Fig. \ref{FIG7}, $\rho_{s}(T)$ does not reach to $T=0$ due to the instability in our numerics. Thus, we extrapolate $\rho_{s}(T)$ up to zero temperature in order to obtain $\rho_{s}(T=0)$.
Other quantity for Homes' law, $\sigma_{DC}(T_{c})$, is determined by \eqref{conduct1} with a numerically computed $T_{c}$. 

\begin{figure}[]
 \centering
     \subfigure[$\rho_{s}/\mu$ at $T=0$]
     {\includegraphics[width=4.75cm]{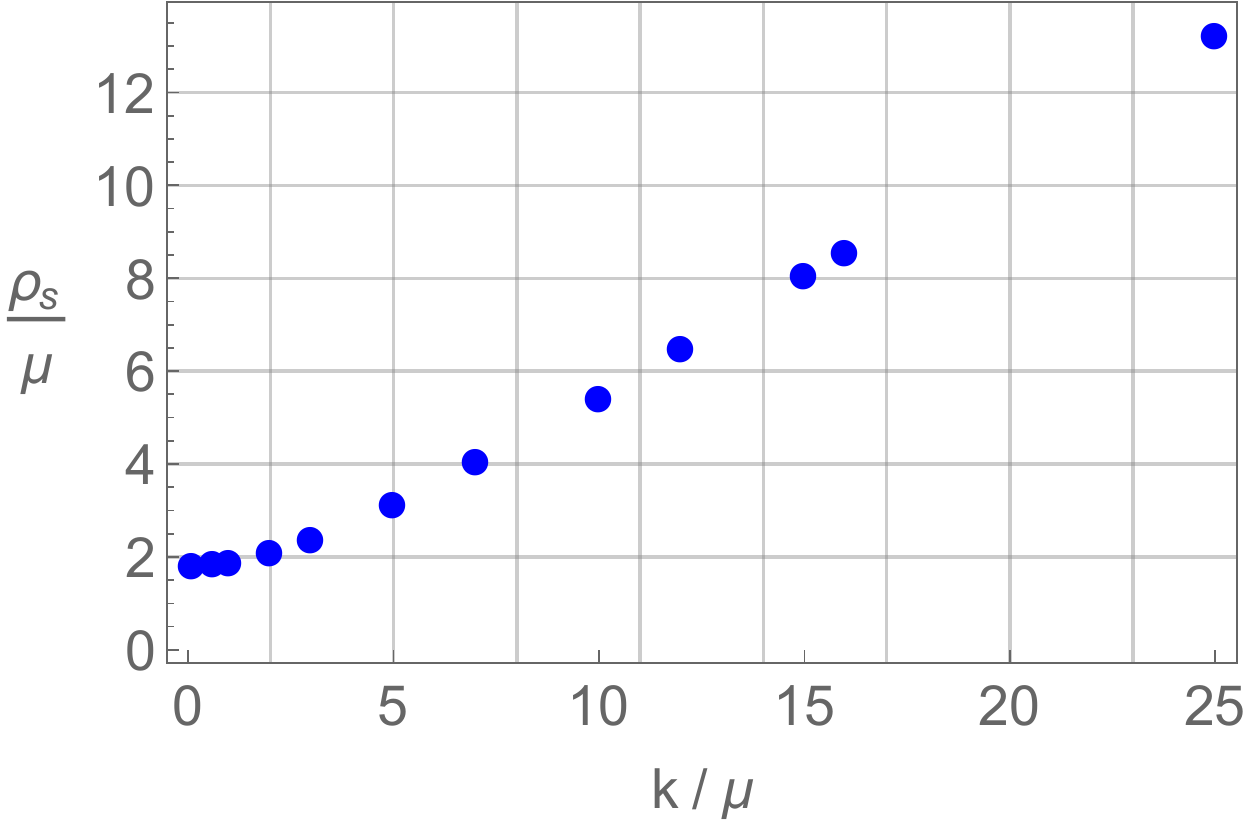} \label{SCALINGTAU1}}
     \subfigure[$T_{c}/\mu$]
     {\includegraphics[width=4.75cm]{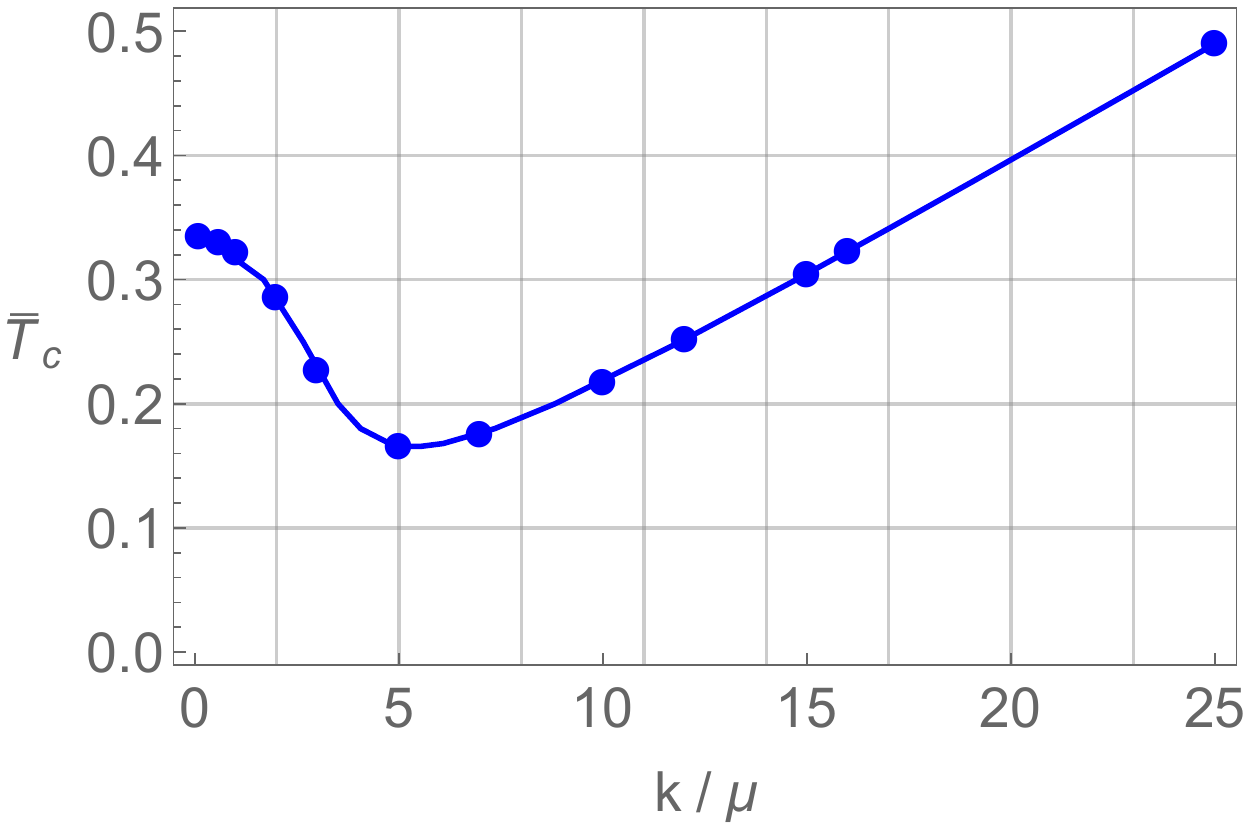} \label{SCALINGTAU2}} 
     \subfigure[$\sigma_{DC}$]
     {\includegraphics[width=4.99cm]{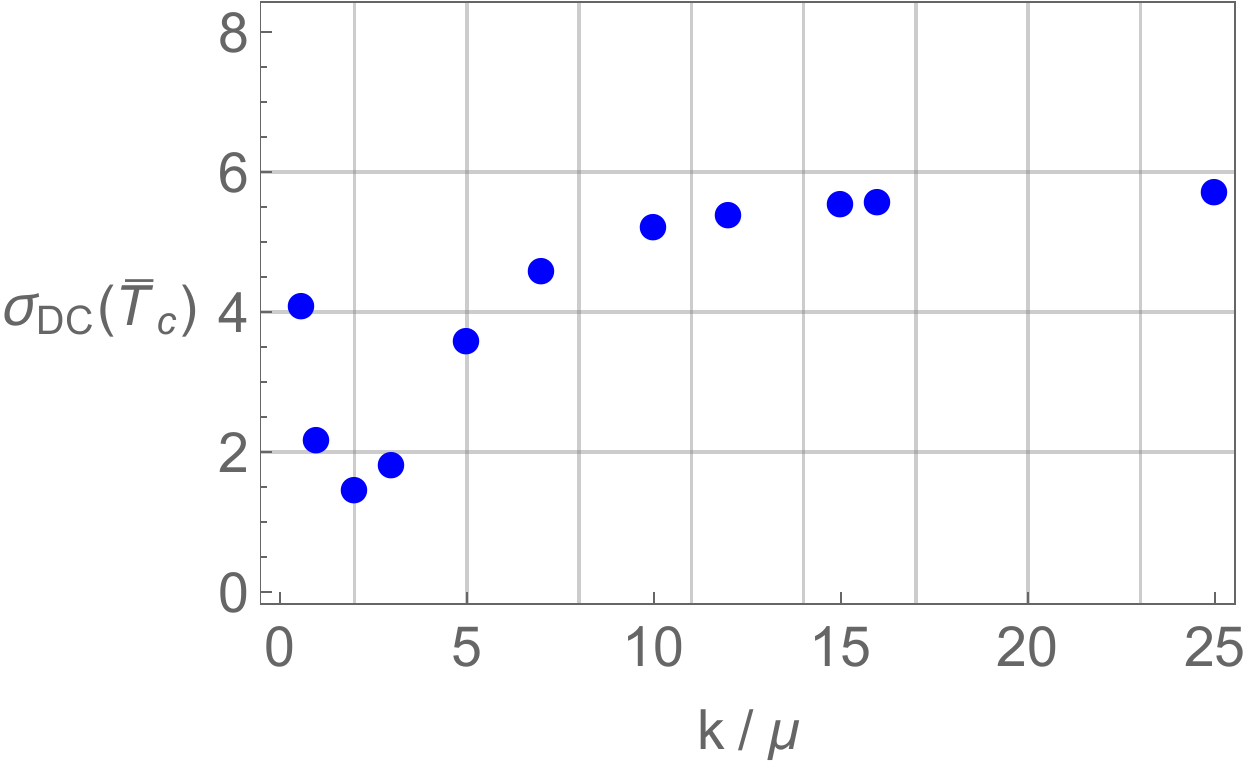} \label{SCALINGTAU3}}
          \caption{$\rho_{s}, T_{c},$ and $\sigma_{DC}(T_{c})$ for $\tau=1/\sqrt{3}$. $\bar{T}_{c}$ is the shorthand notation for $T_{c}/\mu$. The blue solid line corresponds to the one in Fig. \ref{TCFIG}.} \label{SCALINGTAU}
\end{figure}

We first study Homes' law for the scaling case ($\tau=1/\sqrt{3}$). In Fig. \ref{SCALINGTAU}, we display ($\rho_{s}(T=0)$, $T_{c}$, $\sigma_{DC}(T_{c})$).
As we increase $k/\mu$, in the strong momentum relaxation regime, one can see that $\rho_{s}$ and $T_{c}$ are increasing linearly while $\sigma_{DC}$ saturates to some constant.\footnote{$\sigma_{DC}$ tends to diverge in small $k/\mu$ region. This implies the infinite DC conductivity for the weak momentum relaxation.}

Unlike the behavior of $\rho_{s}$ and $\sigma_{DC}$, the increasing behavior of $T_c$ is the distinct property not observed in other holographic studies, for instance, $T_{c}$ tends to decrease or saturates to some value in~\cite{Erdmenger:2015qqa,Kim:2016jjk,Kim:2016hzi}.\footnote{In Fig. \ref{SCALINGTAU2}, the blue line corresponds to the one in Fig. \ref{TCFIG}. This implies that $T_c$ obtained in the fully back-reacted background geometry is consistent with the one without the back-reaction.}  Moreover, this increasing feature from $T_c$ seems to play an important role for Homes' law as we show in shortly.

\paragraph{Homes' law with linear-$T$ resistivity:} 

\begin{figure}[]
 \centering
\subfigure[$\sigma_{DC} T_{c}$ vs $k/\mu$]
     {\includegraphics[width=7.4cm]{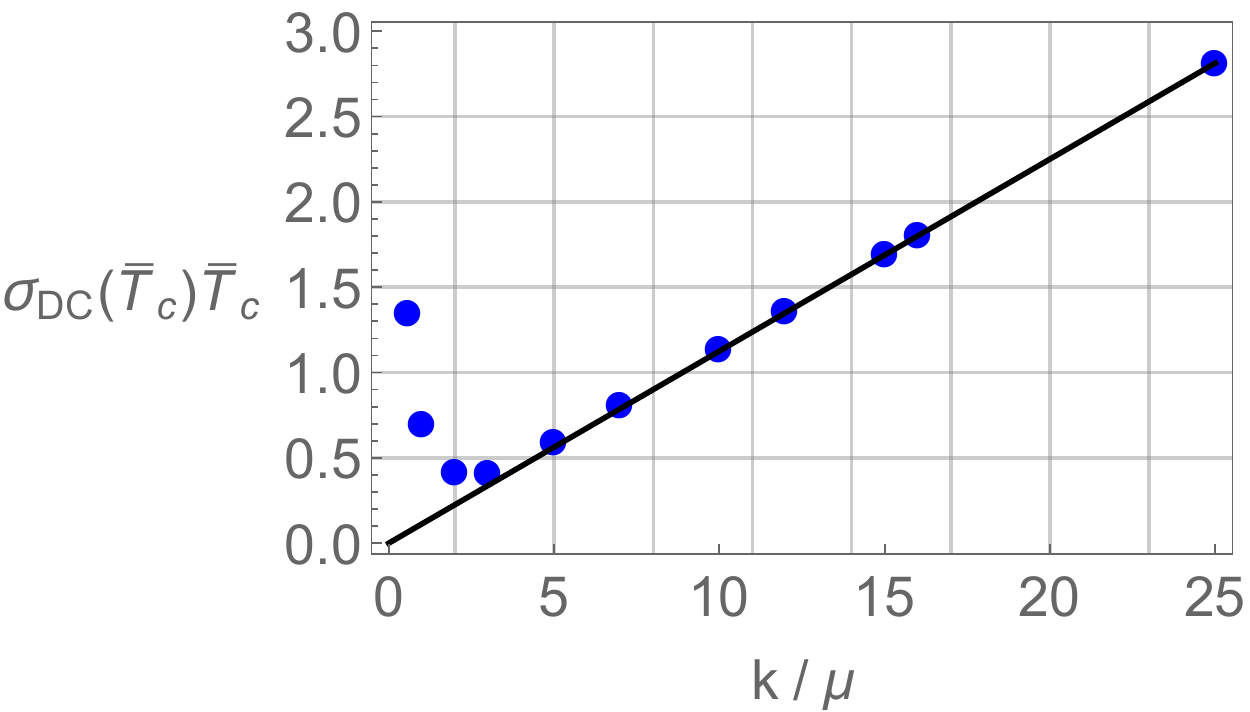} \label{HOMEFIG1}} 
\subfigure[Homes' law]
     {\includegraphics[width=6.8cm]{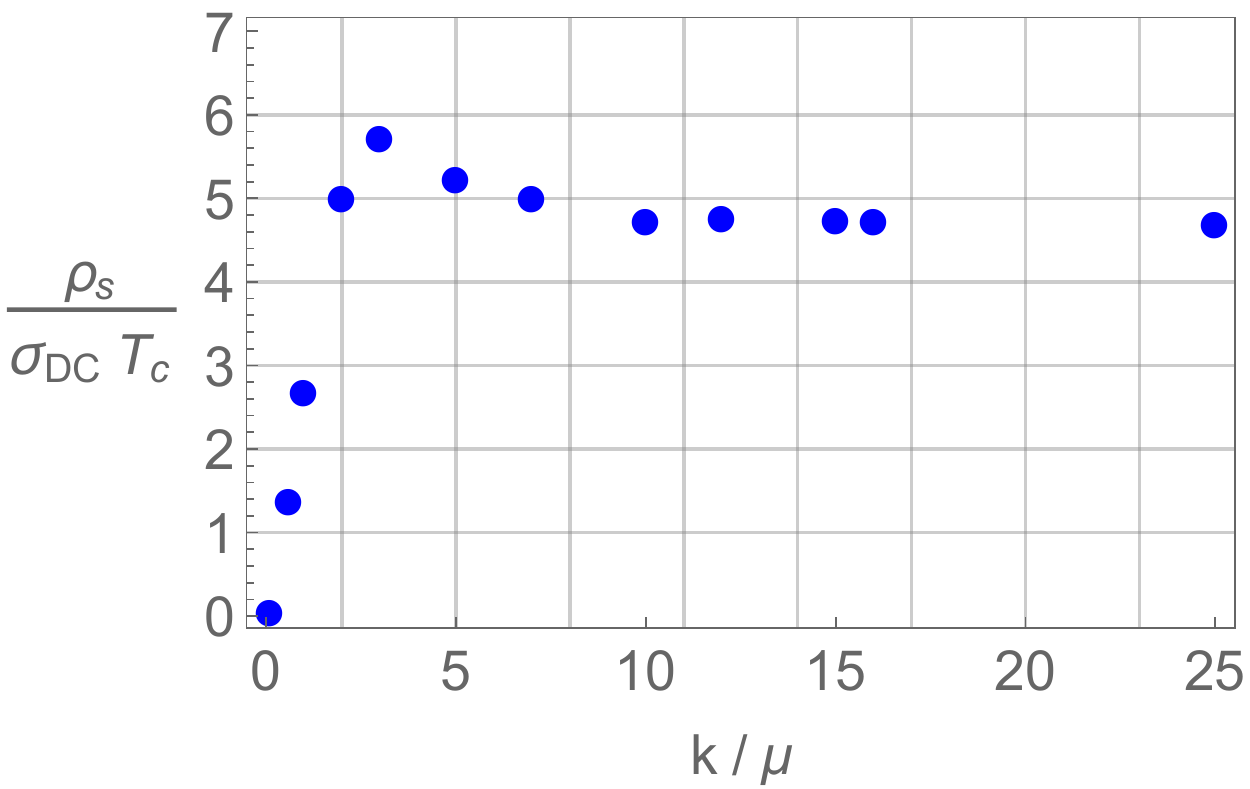} \label{HOMEFIG2}}
          \caption{\textbf{Left:} $\sigma_{DC} T_{c}$ vs $k$. $\bar{T}_{c}$ is the shorthand notation for $T_{c}/\mu$. The black solid line is \eqref{LINEARLINEAR}. \textbf{Right:} Checking Homes' law: Homes' law appears to hold in the strong momentum relaxation limit.} \label{HOMEFIG}
\end{figure}

Using the data in Fig. \ref{SCALINGTAU}, we make a plot of the ratio $\rho_{s}/(\sigma_{DC} T_{c})$ ($=:C$) as a function of $k/\mu$ in Fig. \ref{HOMEFIG2}, and observe that the ratio $C$ becomes a constant, $C \sim 4.7$, at $k/\mu\gg1$ limit, i.e., Homes' law \eqref{HOMEHOME} appears to hold in the strong momentum relaxation limit.
One may wonder if $C$ remains a constant at $k/\mu>25$ in Fig. \ref{HOMEFIG2}. Evaluating $C$ at $\mu/k=0$, we confirmed that $C$ is a constant around $4.7$ in $k/\mu\gg1$ limit.

Homes' law at strong momentum relaxation limit, the constant $C$, can be viewed as the cancelation of the two linearities in $k/\mu$: one from $\rho_{s}$ in Fig. \ref{SCALINGTAU1} and the other from $\sigma_{DC} T_{c}$ in Fig. \ref{HOMEFIG1}. The linearity in Fig. \ref{HOMEFIG1}, the black solid line, can be understood by 
\begin{align}
\sigma_{DC} \, \frac{T}{\mu} \,=\, \frac{1}{2\sqrt{2}\pi} \,\frac{k}{\mu} \,,\label{LINEARLINEAR} 
\end{align}
where the DC conductivity formula \eqref{sigmabigb} is used.
Note that linear-$T$ resistivity in \eqref{sigmabigb} plays a crucial role because  \eqref{LINEARLINEAR} becomes $T$-independent so valid  at $T \gtrsim T_c$.
Alternatively, the linearity in Fig. \ref{HOMEFIG1} (or \eqref{LINEARLINEAR}) may be understood from the fact that $T_c$ is linear in $k$ (Fig. \ref{SCALINGTAU2}) and $\sigma_{DC}$ is constant (Fig. \ref{SCALINGTAU3}).


\subsection{The coupling $\tau$ dependence}\label{taudhw}

Next, let us discuss the $\tau$ dependence in Homes' law.\footnote{In this paper, we perform the computation up to $\tau=\frac{1}{\sqrt{3}}\frac{12}{10}$ because of the stability in our numerics, which might be enough to discuss Homes' law.}
We display the plots for ($\rho_{s}(T=0)$, $T_{c}$, $\sigma_{DC}(T_{c})$) with various $\tau$ in Fig. \ref{SCALINGTAU22}:
\begin{figure}[]
 \centering
     \subfigure[$\rho_{s}/\mu$ at $T=0$]
     {\includegraphics[width =4.70cm]{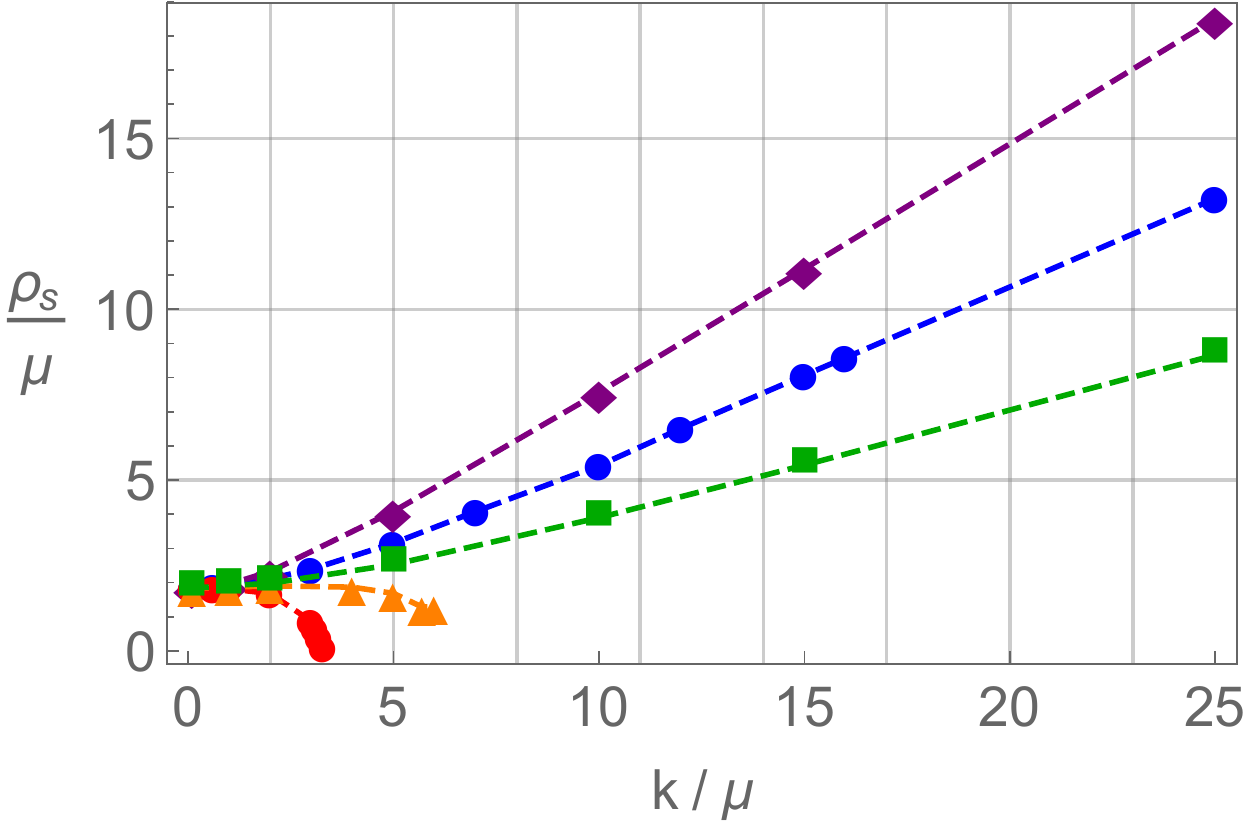} \label{SCALINGTAU22a}}
     \subfigure[$T_{c}/\mu$]
     {\includegraphics[width=4.70cm]{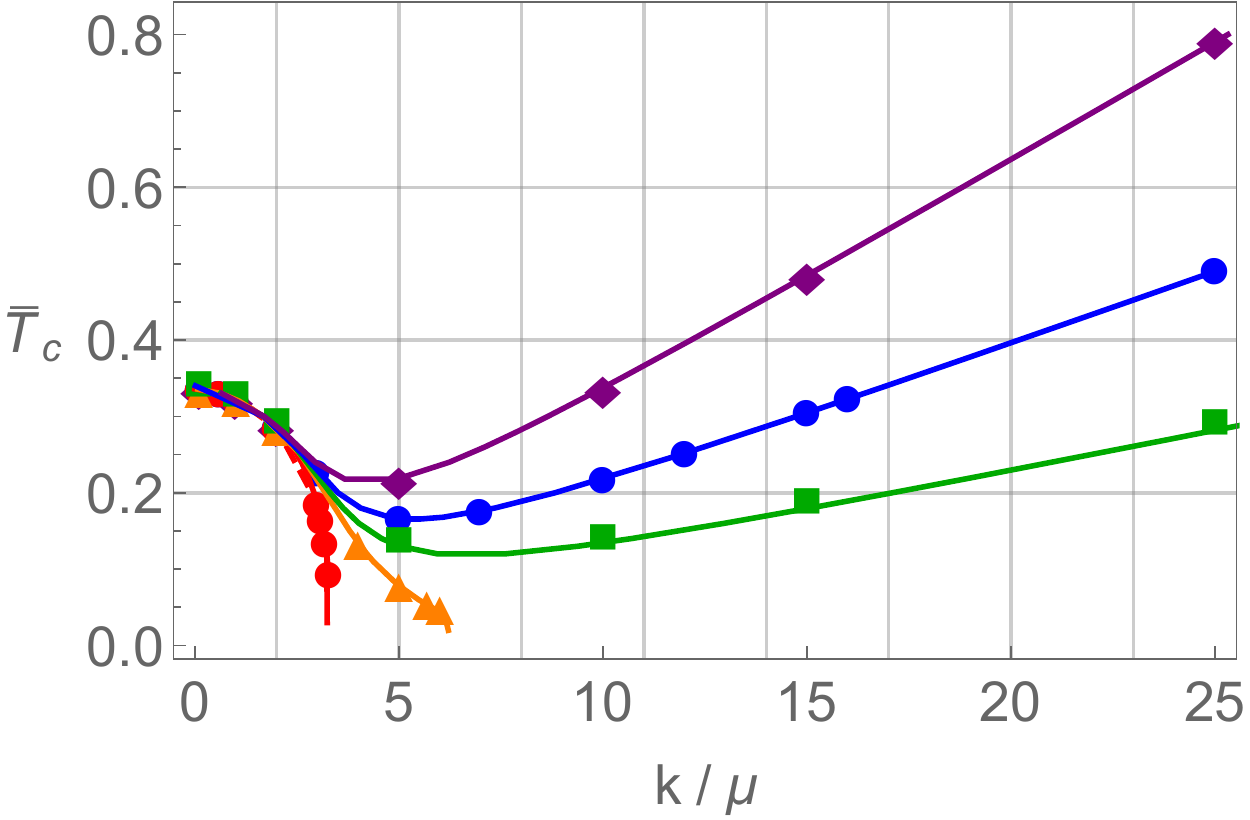} \label{SCALINGTAU22b}} 
     \subfigure[$\sigma_{DC}$]
     {\includegraphics[width=5.06cm]{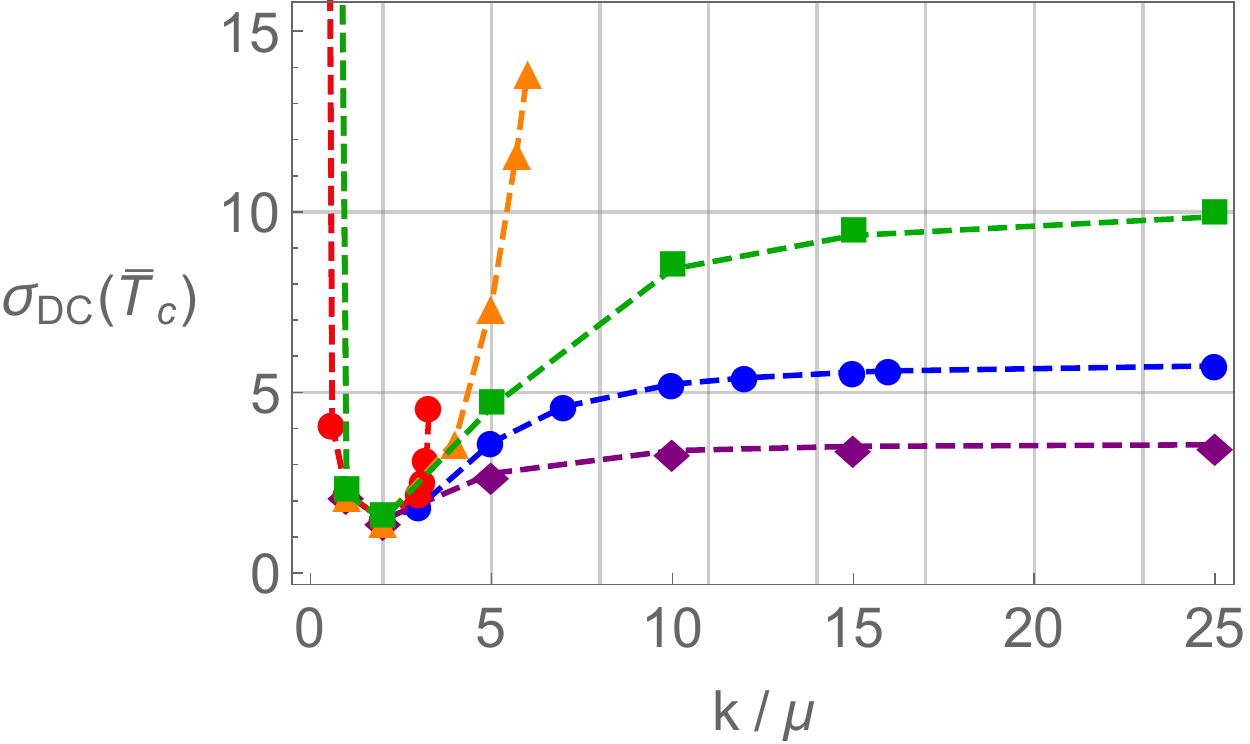} \label{SCALINGTAU22c}}
          \caption{$\rho_{s}/\mu, T_{c}/\mu,$ and $\sigma_{DC}$ for $\tau = (0, \, \frac{1}{\sqrt{3}}\frac{8}{10}, \,\frac{1}{\sqrt{3}}\frac{9}{10}, \,\frac{1}{\sqrt{3}}\frac{10}{10}, \,\frac{1}{\sqrt{3}}\frac{12}{10})$ (red, orange, green, blue, purple). $\bar{T}_{c}$ is the shorthand notation for $T_{c}/\mu$. The solid lines in (b) are Fig. \ref{TCFIG} and the dashed lines in (a), (c) are fitting curves.} \label{SCALINGTAU22}
\end{figure}
blue dots in Fig. \ref{SCALINGTAU22} correspond to Fig. \ref{SCALINGTAU} (i.e., the scaling case $\tau=1/\sqrt{3}$).

\paragraph{$\rho_{s}$ and $T_{c}$ with $\tau$:}
$\rho_{s}(T=0)$ in Fig. \ref{SCALINGTAU22a} shows the qualitatively similar behavior with $T_{c}$ in Fig. \ref{SCALINGTAU22b}: as we increase $k/\mu$, at $\tau<\tau_c$ (red, orange), it is reduced while, at $\tau>\tau_c$ (green, blue, purple), it is linearly increasing.\footnote{This would be another universal relation similar to Homes' law, called Uemura's law which holds only for underdoped cuprates~\cite{Homes:2005aa,Homes:2004wv}: $\rho_s(T=0)=\tilde{C} \, T_c$ with a universal constant $\tilde{C}$.}
One may understand the resemblance between $\rho_{s}(T=0)$ and $T_{c}$ as follows. 
As can be seen in Fig. \ref{FIG7}, $\rho_s(T)$ is zero at $T=T_c$ and monotonically increasing at $T<T_c$. Thus, in order to have a large (small) $\rho_s$ at $T=0$, $T_c$ may need to be large (small) as well, i.e., $\rho_s (T=0) \sim T_{c}$.

Note that the solid lines in Fig. \ref{SCALINGTAU22b} are Fig. \ref{TCFIG}, i.e., $T_c$ from the fully back-reacted geometry is consistent with the one without back-reaction. This may imply that $\rho_s$ might also be understood in the probe limit. If one can develop the methodology to compute $\sigma(\omega)$ only with the IR geometry (the $T=0$ analysis), we suspect that the numerical result in Fig. \ref{SCALINGTAU22a} might also be confirmed in a simple probe limit with the scaling property \eqref{SAS}.

\paragraph{$\sigma_{DC}(T_{c})$ with $\tau$:}
In Fig. \ref{SCALINGTAU22c}, unlike $\rho_{s}$, there is no resemblance between $\sigma_{DC}(T_{c})$ and $T_{c}$. At small $k/\mu$, $\sigma_{DC}$ is diverging independent of $\tau$, which reflects the fact the conductivity is infinite at zero momentum relaxation.

On the other hand, at larger $k/\mu$, one can see the $\tau$ dependence on $\sigma_{DC}(T_c)$.
At $\tau<\tau_c$ (red, orange), it is increasing while, at $\tau>\tau_c$ (green, blue, purple), it saturates to some constant.
Note that the behavior of $\sigma_{DC}(T_c)$ at large $k/\mu$ depends on the holographic models: for instance, it is increasing in the Q-lattice model~\cite{Kim:2016jjk} or saturated in the linear axion model~\cite{Kim:2016hzi}.

\begin{figure}[]
 \centering
\subfigure[$\sigma_{DC} T_{c}$ vs $k/\mu$]
     {\includegraphics[width=7.4cm]{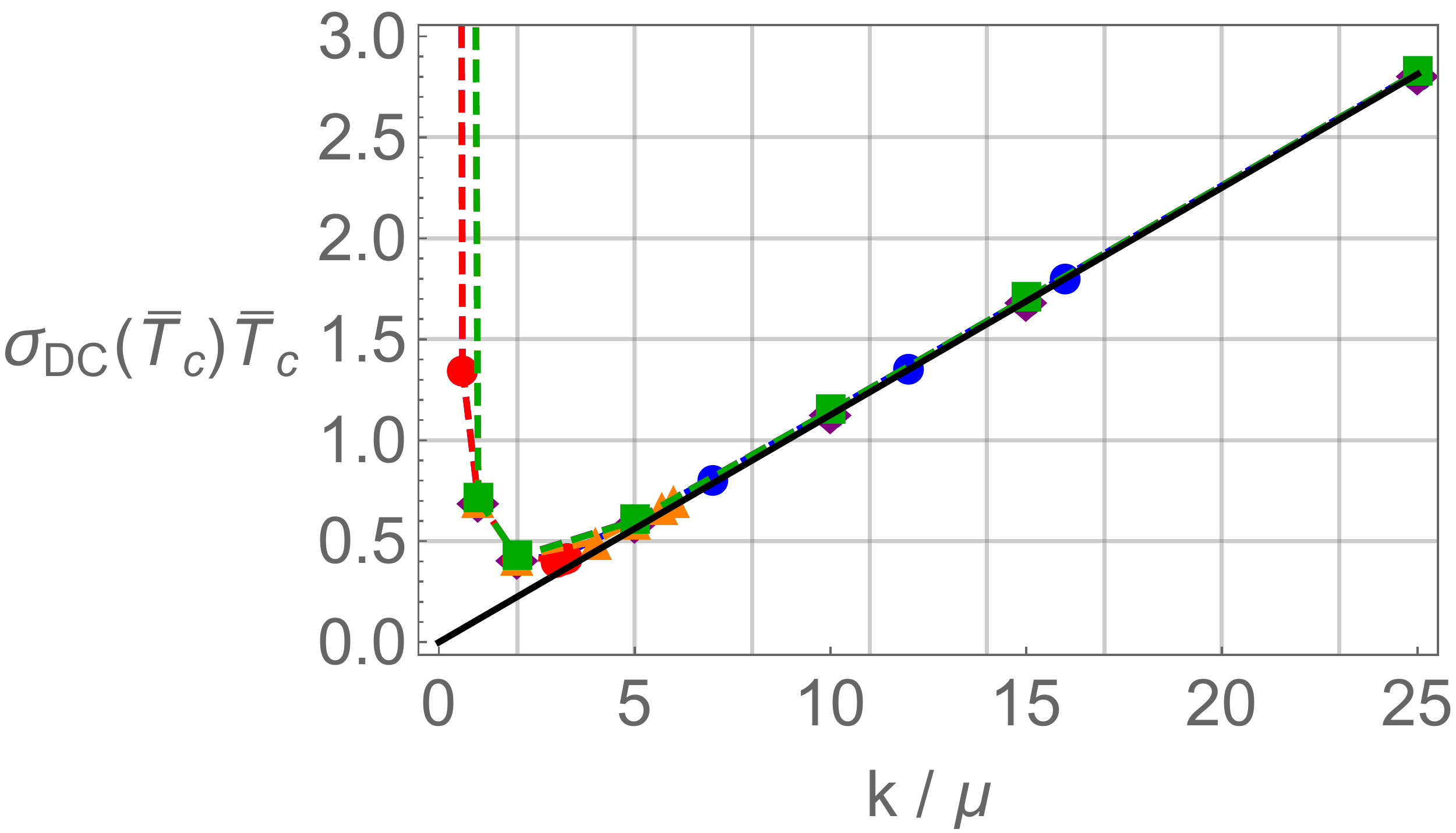} \label{HOMEFIGFIG1}} 
\subfigure[Homes' law]
     {\includegraphics[width=6.8cm]{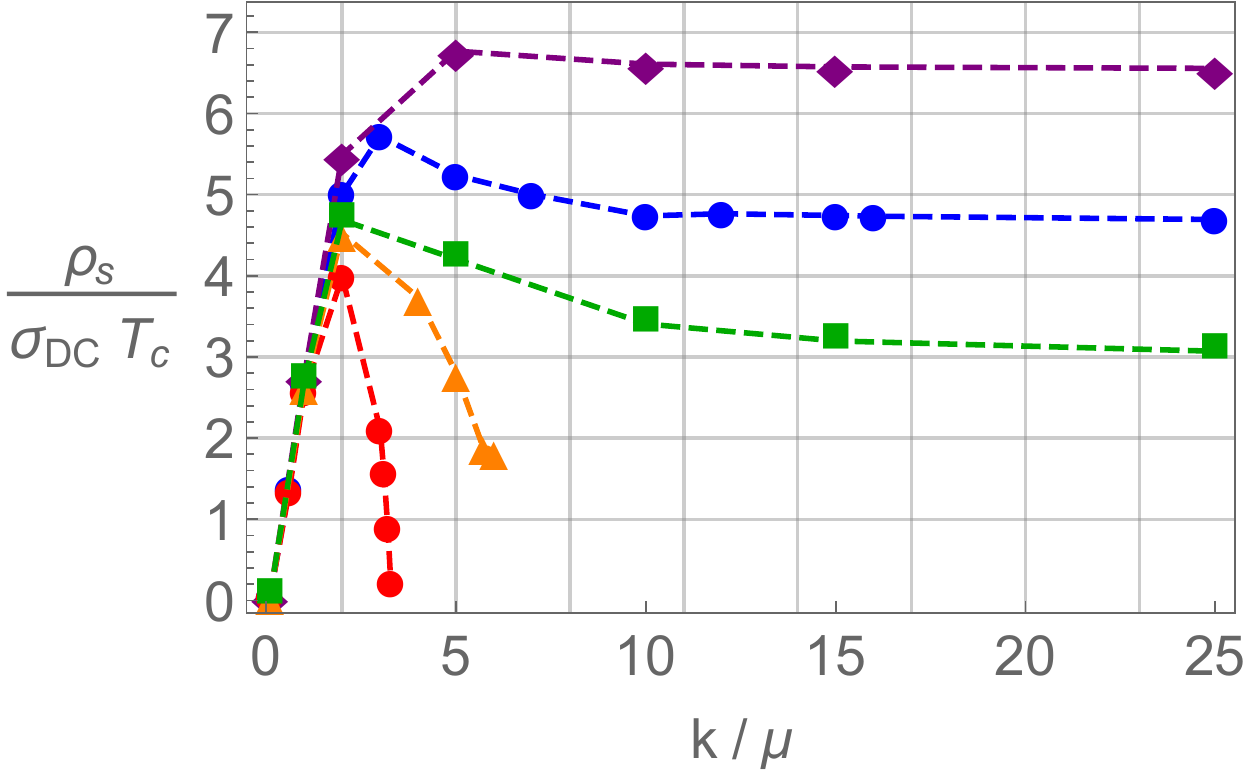} \label{HOMEFIGFIG2}}
          \caption{\textbf{Left:} $\sigma_{DC} T_{c}$ vs $k$. The black solid line is \eqref{LINEARLINEAR}. \textbf{Right:} Checking Homes' law with $\tau = (0, \, \frac{1}{\sqrt{3}}\frac{8}{10}, \,\frac{1}{\sqrt{3}}\frac{9}{10}, \,\frac{1}{\sqrt{3}}\frac{10}{10}, \,\frac{1}{\sqrt{3}}\frac{12}{10})$ (red, orange, green, blue, purple).} \label{HOMEFIGFIG}
\end{figure}

\paragraph{Homes' law at $\tau>\tau_{c}$:}
With the data in Fig. \ref{SCALINGTAU22}, we find Homes' law \eqref{HOMEHOME} at the strong momentum relaxation limit for $\tau>\tau_{c}$ (green, blue, purple): i.e., $C$ become a constant at $k/\mu\gg1$. See Fig. \ref{HOMEFIGFIG2}.
As did in the scaling case ($\tau=1/\sqrt{3}$) in Fig. \ref{HOMEFIG1}, Homes' law at $\tau>\tau_{c}$ can be understood from the cancelation of the two linearities in $k/\mu$: one from $\rho_{s}$  (Fig. \ref{SCALINGTAU22a}) and the other from the linear-$T$ resistivity (Fig. \ref{HOMEFIGFIG1}). 

We also find that the saturating value of $C$ depends on the value of $\tau$. Thus, $\tau$ might be used to match the experiment data: $C=4.4$ for ab-plane high $T_{c}$ superconductors as well as clean BCS superconductors and $C=8.1$ for c-axis high $T_{c}$ superconductor and BCS superconductors in the dirty limit.

\section{Conclusions} \label{sec4}
We have investigated Homes' law \eqref{HOMEHOME} by computing the critical temperature $T_c$, the superfluid density $\rho_s$ at zero temperature, and the DC conductivity $\sigma_{DC}$ at $T_c$ in a holographic superconductor based on the Gubser-Rocha model \eqref{action1} with the minimally chosen coupling term $B(\phi)$:
\begin{align} \label{CSBDE223}
B(\phi) = M^{2} \, \cosh \left( \tau \phi \right) \,,
\end{align}
where it corresponds to the mass term of the complex scalar field, $M^2$, at $\tau=0$.
The action \eqref{action1} also contains the axion field to study the momentum relaxation where its strength is denoted as $k/\mu$.
In this setup, Homes' law means that $C:=\rho_s/(\sigma_{DC}T_c)$ is independent of the momentum relaxation. 

The Gubser-Rocha model, a \textit{normal} phase, is appealing in that the linear-$T$ resistivity can be obtained at strong momentum relaxation limit ($k/\mu\gg1$) above $T_c$. Considering the complex scalar field with the Gubser-Rocha model, we find that, in order to study the \textit{superconducting} phase at $k/\mu\gg1$, $\tau$ in the coupling \eqref{CSBDE223} is important. We show that the  conditions to study a holographic superconductor having the linear-$T$ resistivity above $T_c$ would be:
\begin{align}\label{clfew22}
\text{i)} \,\, \tau>\tau_{c};   \qquad\qquad  \text{ii)} \,\,  k/\mu\gg1 \,,
\end{align}
where $\tau_c\neq0$ can be determined numerically from the instability analysis for $T_c$.  {The first condition i) in \eqref{clfew22} means that if $\tau$ is smaller than $\tau_c$ the superconducting phase does not exist at large $k/\mu$.  In particular, }the trivial coupling term $B(\phi)=M^2$ ($\tau=0$ case), the mass term of the complex scalar field, used in previous literature can not capture a complete feature of the superconducting phase at $k/\mu\gg1$. 

With the condition \eqref{clfew22}, we find Homes' law can hold in the strong momentum relaxation limit, i.e., $C$ becomes a constant at $k/\mu\gg1$ limit.
In \cite{Hartnoll:2014lpa}, it is argued that if the momentum is relaxed {\it quickly(strongly)}, which is an extrinsic so non-universal effect, transport can be governed by an intrinsic and universal effect such as diffusion of energy and charge. Thus, the universality of linear-$T$ resistivity may appear in the regime of {\it strong} momentum relaxation (so called the {\it incoherent} regime). 
 Consequently, Homes' law can appear also in the strong momentum relaxation limit.

Furthermore, we showed that Homes' law at $k/\mu\gg1$ can be understood from the cancelation of the two linearities in $k/\mu$: one from $\rho_{s}(T=0)$ (numerical result) and the other from the linear-$T$ resistivity (analytic result) \eqref{LINEARLINEAR}.
It will be interesting to show the linearity of $\rho_{s}(T=0)$ in $k/\mu$ also analytically. If one can develop the method to compute the optical conductivity with the IR geometry ($T=0$ analysis) in the superconducting phase\footnote{In \cite{Horowitz:2009ij,Basu:2011np,Yang:2019gce}, the analytic IR geometry is given for $M^2=0, \tau=0$ and $k/\mu=0$.}, we suspect that the linearity of $\rho_{s}(T=0)$  in $k/\mu$ might be related with the IR scaling property of the coupling $B(\phi)$ \eqref{SAS}.

We find that the value of $C$ at $k/\mu\gg1$ depends on the value of $\tau$ so $\tau$  can  be used to match the experiment data: $C=4.4$ for ab-plane high $T_{c}$ superconductors as well as clean BCS superconductors and $C=8.1$ for c-axis high $T_{c}$ superconductor and BCS superconductors in the dirty limit.

It may also be interesting to study Homes' law with the holographic models having other IR geometries~\cite{Gouteraux:2014hca,Ahn:2017kvc,Ahn:2019lrh} {together with linear $T$ resistivity.}
In \cite{Ahn:2019lrh}, authors found that when the IR geometry is governed by a finite dynamical exponent $z$ and a hyperscaling violating exponent $\theta$ unlike the Gubser-Rocha model ($z\rightarrow\infty$, $\theta\rightarrow-\infty$), the linear-$T$ resistivity can also exhibit at high temperature if the momentum relaxation is strong. 
{Therefore, one may investigate how much general our results in this paper are. One may also check if $\tau>\tau_{c}$ condition in \eqref{clfew22} is necessary for Homes' law in more generic setup.}
We leave these subjects as future work and hope to address them in the near future.

\acknowledgments

We would like to thank  {Yongjun Ahn}  for valuable discussions and correspondence.  
This work was supported by the National Key R$\&$D Program of China (Grant No. 2018FYA0305800), Project 12035016 supported by National Natural Science Foundation of China, the Strategic Priority Research Program of Chinese Academy of Sciences, Grant No. XDB28000000, Basic Science Research Program through the National Research Foundation of Korea (NRF) funded by the Ministry of Science, ICT $\&$ Future Planning (NRF- 2021R1A2C1006791) and GIST Research Institute(GRI) grant funded by the GIST in 2021.

\bibliographystyle{JHEP}

\providecommand{\href}[2]{#2}\begingroup\raggedright\endgroup

\end{document}